\documentclass[aps,prd,preprint,floats,epsf,superscriptaddress,nofootinbib]{revtex4}

\usepackage{graphicx} % Packages to allow inclusion of graphics
\usepackage{amsmath,amssymb,amsfonts,mathtools} % Typical maths resource packages
\usepackage{mathrsfs}
\usepackage{slashed} % Feynman slashes
\usepackage{braket} % Bracket notation for quantum physics
\usepackage{indentfirst}
\usepackage{xcolor}
\usepackage{dsfont}
\usepackage{ulem}
\usepackage{dcolumn}% Align table columns on decimal point
\usepackage{bm}% bold math
\usepackage{bbm}% bold math for {\mathbbm 1}

\usepackage[T1]{fontenc}
\usepackage{calligra}
\usepackage[T1]{pbsi}
\usepackage{CJKutf8} % Chinese typing

\usepackage[colorlinks,
            linkcolor=black,
            anchorcolor=black,
            citecolor=black
            ]{hyperref} % For creating hyperlinks in cross references

\begin{document} 

\title{Probing dark gauge boson with observations from neutron stars}

\author{Bo-Qiang Lu}
\email[e-mail: ]{bqlu@zjhu.edu.cn}
\affiliation{School of Science, Huzhou University, Huzhou, Zhejiang 313000, China}
\author{Cheng-Wei Chiang}
\email[e-mail: ]{chengwei@phys.ntu.edu.tw}
\affiliation{Department of Physics and Center for Theoretical Physics, National Taiwan University, 
Taipei, Taiwan 10617, Republic of China}
\affiliation{Physics Division, National Center for Theoretical Sciences, Taipei, Taiwan 10617, Republic of China}

\begin{abstract}
We present an investigation on the production of light dark gauge bosons by the nucleon bremsstrahlung processes in the core of neutron stars. 
The dark vector is assumed to be a $U(1)_{B-L}$ gauge boson with a mass much below keV. 
We calculate the emission rate of the dark vector produced by the nucleon bremsstrahlung in the degenerate nuclear matter. 
In addition, we take into account the photon-dark vector conversion for the photon luminosity observed at infinity.
Combining with the observation of J1856 surface luminosity, we find that a recently discovered excess of J1856 hard x-ray emission in the 2–8~keV energy range 
by XMM-Newton and Chandra x-ray telescopes could be consistently explained by a dark vector with gauge coupling $e^{\prime}=5.56\times 10^{-15}$, 
mixing angle $\varepsilon=1.29\times 10^{-9}$, and mass $m_{\gamma^{\prime}}\lesssim 10^{-5}$~eV. 
We also show that the mixing angle $\varepsilon > 7.97 \times 10^{-9}$ for $m_{\gamma^{\prime}} \lesssim 3 \times 10^{-5}~\rm eV$ and the gauge 
coupling $e^{\prime} > 4.13 \times 10^{-13}$ for $m_{\gamma^{\prime}} \lesssim 1~\rm keV$ have 
been excluded at 95\% confidence level by the J1856 surface luminosity observation.
Our best-fit dark vector model satisfies the current limits on hard x-ray intensities from the Swift and INTEGRAL hard x-ray surveys.
Future hard x-ray experiments such as NuSTAR may give a further test on our model.

\end{abstract}

\maketitle
%\flushbottom

%%%%%%%%%%%%%%%%%%%%%%%%%%%%%%%%%%%%%%%%%%%%%%%%%%
\section{Introduction}
\label{sec:intro}
%%%%%%%%%%%%%%%%%%%%%%%%%%%%%%%%%%%%%%%%%%%%%%%%%%

One of the significant puzzles in particle physics that demand new theory beyond the Standard Model (SM) is the nature of dark matter (DM), whose existence has been confirmed from various cosmological and astrophysical observations~\cite{Planck2016AA}.  
The most popular class of DM candidates is the weakly interacting massive particles (WIMPs)~\cite{Lee1977PRL, Hut1977PLB} with a mass at around 
the electroweak scale and a coupling strength similar to the weak coupling. However, the WIMP hypothesis has suffered stringent constraints 
due to the null results from both direct~\cite{XENON1T2018} and indirect~\cite{Ackermann2015PRL} DM detection experiments.
As an alternative candidate of DM, a hidden sector consisting of very low-mass particles has drawn more attention in recent years.
One of the simplest extensions of the SM is to include an additional $U(1)$ gauge group~\cite{Holdom1986PLB,Chiang2020PRD}, which can naturally arise from the breaking of grand 
unified theory (GUT) groups~\cite{Dienes1997NPB,Ibe2019JHEP} or the string theory~\cite{Dienes1997NPB,Lukas2000JHEP,Blumenhagen2005JHEP,Abel2008JHEP,
Burgess2008JHEP,Goodsell2009JHEP,Burrage2009JCAP,Cicoli2011JHEP}. 
The gauge boson associated with the new $U(1)$ group, dubbed the dark photon (which mixes with the SM photon), may play the role of a mediator 
between the SM and dark sectors or as a dark matter candidate~\cite{Fabbrichesi2020}.

Supernovae serve as a powerful factory for the emission of dark photons with masses $\lesssim 100$~MeV~
\cite{Bjorken1009PRD,Dent2012,Kazanas2014NPB,Rrapaj2016PRC,Chang2017JHEP,Hardy2017JHEP,DeRocco2019JHEP,Hong2020}. With a sufficiently weak 
interaction, dark photons generated within the core of a supernova can escape the progenitor star before their decays, and will contribute to 
the stellar energy transport.  The supernova cooling would be accelerated if the dark photons could transport energy out of the core more 
efficiently than the standard supernovae cooling via the emission of neutrinos.  The bounds on dark photons by using the measurements of 
SN 1987a have recently been updated in Refs.~\cite{Chang2017JHEP,Hardy2017JHEP,DeRocco2019JHEP}, with the consideration of plasma effects for the production of dark photons.
The Sun, horizontal branch (HB) stars, and red giants, on the other hand, can serve as important laboratories for dark photons in the mass range of $\lesssim $~keV.
For instance, Refs.~\cite{An2013PLB,An2013PRL} show that the measured luminosity of the Sun, $L_{\odot}=3.83\times 10^{26}$~W, has constrained the dark photon parameter space to the range of $\varepsilon m_{\gamma'}<4\times 10^{-12}~{\rm eV}$~\cite{An2013PRL}.
% These bounds are often referred to as `stellar cooling', though they more properly correspond to anomalous energy loss or transport.

In this work, we will focus on the production of dark gauge bosons in the core of neutron stars (NSs), which are the remnants of the supernova 
explosion and have long been recognized as excellent laboratories to search for axionlike 
particles~\cite{Iwamoto1984PRL,Morris1986PRD,Brinkmann1988PRD,Iwamoto2001PRD,Lai2006PRD,Fortin2018JHEP,Sedrakian2016PRD,Sedrakian2019PRD,
Carenza2019JCAP,Harris2020JCAP,Buschmann2021PRL,Leinson2019JCAP,Leinson2021JCAP,Fortin2021}.
The core of a NS is composed of highly degenerate nuclear matter that has an average density of $\sim 3\times 10^{14}$~$\rm g/cm^3$, 
which corresponds to an average distance of $\sim 1$~fm between nucleons.
Dark gauge bosons can be produced through bremsstrahlung of proton and neutron scattering in the NS cores.
In the previous literature~\cite{Chang2017JHEP,Hardy2017JHEP,DeRocco2019JHEP}, the nondegenerate limit has been assumed to obtain the dark 
vector emission rate as well as the constraints on the dark vector from the supernova, where the plasma is thought to be diluted and nonrelativistic. 
However, the nuclear matter in the NS core is in fact highly compressed, and therefore, in this work, we will present the calculation of the production of 
the dark vector in the NS core in the degenerate limit.

The in-medium effects can lead to a suppression on the dark vector production rate if the dark vector which couples coherently to the 
charged SM plasma has a mass much below the stellar plasma frequency~\cite{Chang2017JHEP,Hardy2017JHEP}.
In additional to the electromagnetic (EM) current through mixing, the new gauge boson could couple to the $B-L$ current in the well-known 
theory of $U(1)_{B-L}$ extension of the SM~\cite{Davidson1979PRD,Marshak1980PLB,Mohapatra1980PRL,Davidson1987PRL,Montero2009PLB,Ma2014PLB}.
For such a theory, the new vector boson will dominantly be produced by the neutron bremsstrahlung in the NS core.  
This is because, for one thing, neutron is the dominant component of the NS, and for another, the vector boson emitted by the neutron in the 
initial or final state is not suppressed by the plasma effects.  In this work, we will assume that the dark gauge boson is the $B-L$ gauge boson 
with a mass much below keV, the core temperature of the NS, and the gauge coupling $e'$ is sufficiently small.

Recently, a significant excess of hard x-ray emission in the $2-8$~keV energy range was found by analyzing the data from the 
XMM-Newton and Chandra x-ray telescopes~\cite{Dessert2020APJ} observing on the nearby magnificent seven (M7) x-ray dim isolated NSs. 
The excess was interpreted as the emission of axions by the NS in Ref.~\cite{Buschmann2021PRL} and was found to be consistent with the current constraints.
In this work, we explore the interpretation of a dark vector scenario for the J1856 hard x-ray excess.
We calculate the dark vector emission rate in the NS core and simulate the evolution of the NS based on the modified {\tt NSCool} code~\cite{Page2016}, including the dark vector emissivity. 
Due to the strong magnetic field of the magnetospheres surrounding the NSs, the dark vectors produced in the NS cores may convert into x-rays when they propagate outwards~\cite{Fortin2019JCAP}. We present an analytical formula
of the dark vector-photon conversion probability for dark vector with mass $m_{\gamma^{\prime}}\lesssim (T_s/R_s)^{1/2}$, where $T_s$ and $R_s$ are the surface temperature and radius of the NS, respectively.
We also calculate the surface luminosity observed at infinity by taking into account the photon-dark vector conversion.  Based on the Bayesian inference, the analysis of J1856 hard x-ray data, as well as surface luminosity observation are carried out by the {\tt UltraNest} 
package~\cite{Buchner2021}. We find that the dark vector scenario is favored by the hard x-ray excess, and we also derive 95\% upper limits on the model parameters from the data.

The structure of this paper is arranged as follows.  Section~\ref{sec:The model} defines the dark gauge boson model considered in this work, and 
sets up the required notations.  In Sec.~\ref{sec:dp}, we calculate the production of dark vectors from nucleon bremsstrahlung, including 
the squared matrix element and emission rate.  
In Sec.~\ref{sec:nplum}, we briefly review the neutrino and photon luminosities for the NS cooling. We also calculate the photon surface luminosity 
observed at infinity that includes photon-dark vector conversion.
In Sec.~\ref{sec:coolSimulate}, we describe the NS simulation based on the modified {\tt NSCool} code in detail.
In Sec.~\ref{sec:plum}, we compute the hard x-ray spectrum from the dark vector-photon conversions.
The statistical analysis of data based on the Bayesian inference is carried out in Sec.~\ref{sec:dataAnalysis}, and the 95\% upper limits on the dark vector model 
are derived in Sec.~\ref{sec:limits}. Finally in Sec.~\ref{sec:conclusion}, we summarize our findings.  
The plasma effects on dark vector's production are discussed in Appendix~\ref{adp:imf}.
The dark vector-photon conversion probability is derived in Appendix~\ref{apd:prob}.
In Appendix~\ref{apd:analyticalProb}, we show the analytical results for the dark vector-photon conversion probability. 
The detailed calculations of the dark vector emission rate are given in Appendix~\ref{apd:degenerate}.
Discussions on the one-pion approximation are presented in Appendix~\ref{apd:OPE}.

%%%%%%%%%%%%%%%%%%%%%%%%%%%%%%%%%%%%%%%%%%%%%%%%%%
\section{The model}
\label{sec:The model}
%%%%%%%%%%%%%%%%%%%%%%%%%%%%%%%%%%%%%%%%%%%%%%%%%%

We are interested in the extension of the SM with an additional $U(1)$ gauge symmetry, having a new gauge boson $A_{\mu}^{\rm D}$ called 
the dark vector. The relevant effective Lagrangian at low scales is written as
\begin{equation}
  \label{eq:u1larg}
  \begin{aligned}
  \mathcal{L}_{\mathrm{eff}}=-\frac{1}{4}F_{\mu \nu}^{\prime}F^{\prime \mu \nu}-\frac{1}{2} m_{\gamma^{\prime}}^{2}A_{\mu}^{\prime}A^{\prime\mu}
  +\frac{\varepsilon}{2} F_{\mu \nu} F^{\prime \mu \nu}+e^{\prime} A_{\mu}^{\rm D} J^{\mu},
  \end{aligned}
\end{equation}
where we assume the dark vector has a Stueckelberg mass $m_{\gamma'}$, $F_{\mu \nu}=\partial_{\mu} A_{\nu}^{\rm SM}-\partial_{\nu} A_{\mu}^{\rm SM}$ and 
$F_{\mu \nu}^{\prime}=\partial_{\mu} A_{\nu}^{\rm D}-\partial_{\nu} A_{\mu}^{\rm D}$ are, respectively, the field strengths of the SM photon and dark gauge boson, and
$\varepsilon$ represents the mixing angle between the SM photon and dark vector. The parameter $e'$ denotes the coupling between $A_{\mu}^{\rm D}$ and the current $J^{\mu}$, 
which in this work is assumed to be the SM $B-L$ current arising from a $U(1)_{B-L}$ symmetry, as widely investigated in the 
literature~\cite{Davidson1979PRD,Marshak1980PLB,Mohapatra1980PRL,Davidson1987PRL,Montero2009PLB,Ma2014PLB}.
The kinetic terms of gauge boson in the Eq.~\eqref{eq:u1larg} can be diagonalized by rotating the gauge fields as 
\begin{equation}
    \left(\begin{array}{c}
    A_{\mu}^{\rm D} \\
    A_{\mu}^{\rm SM}
    \end{array}\right)=\frac{1}{\sqrt{1-\varepsilon^2}}
    \left(\begin{array}{cc}
    1 & 0 \\
    -\varepsilon & \sqrt{1-\varepsilon^2}
    \end{array}\right)\left(\begin{array}{cc}
    \cos \theta & -\sin \theta \\
    \sin \theta & \cos \theta
    \end{array}\right)\left(\begin{array}{c}
    A_{\mu}^{\prime} \\
    A_{\mu}
\end{array}\right)
~,
\end{equation}
where $\left(A_{\mu}^{\rm D},~A_{\mu}^{\rm SM} \right)^T$ and $\left(A_{\mu}^{\prime},~A_{\mu} \right)^T$ are the so-called interaction 
eigenstates and mass (propagating) eigenstates, respectively.
For a massive dark vector, the rotation angle $\theta$ is locked at zero~\cite{Fabbrichesi2020}.
We see that the mixing leads to the interactions between the dark vector and the SM charged particles, 
such as electron, proton and charged pions, with a coupling strength $\varepsilon e$. Furthermore, due to the mixing, the dark vector
can convert to a photon during their propagation. 
% In this case, the nearly massless (compared to the keV scale) dark photon $X_{\mu}$ couples to both visible and dark currents while the 
% ordinary photon $A_{\mu}$ couples exclusively to the visible current:

In the plasma of supernovae, the visible photon develops a nonzero mass that is determined by the plasma frequency $\omega_p^2=4\pi \alpha_{\mathrm{EM}} \sum_i n_{i}/E_{F,i}$, 
where the sum goes over the charged particles with a number density $n_i$ and a Fermi energy $E_{F,i}^{2}=m_{i}^{2}+\left(3 \pi^{2}n_{i}\right)^{2/3}$.
The nonzero plasma mass of the photon can lead to inequivalent effective couplings to the charged particles in the plasma and in the vacuum 
(the Lagrangian parameter). The effective coupling between the SM fermions and $A_{\mu}^{\prime}$ in the medium is given by~\cite{Hong2020}
\begin{equation}
  \label{eq:eeff}
  e_{\mathrm{eff}}^{f}=e^{\prime}\left(q_{e}^{\prime} q_{f}-q_{f}^{\prime} q_{e}\right)+\left(\varepsilon e-q_{e}^{\prime} 
  e^{\prime}\right) q_{f} \frac{m_{\gamma^{\prime}}^{2}}{m_{\gamma^{\prime}}^{2}-\pi_{T,L}},
\end{equation}
where $f$ denotes the SM fermions with an electric charge $q_f$ and a dark gauge quantum number $q_f^{\prime}$
in units of $e$ and $e'$, respectively. The proton and neutron have the electric charges $q_p=1$ and $q_n=0$, and the dark gauge quantum numbers 
$q_p^{\prime}=q_n^{\prime}=1$. The EM and dark gauge quantum numbers for the electron are $q_e = q_e^{\prime}=-1$. 
The visible photon mass is encoded in the polarization tensor $\pi_{T,L}$ given in Appendix~\ref{adp:imf}.

From Eq.~\eqref{eq:eeff}, the effective couplings of electron, proton, and neutron to $A_{\mu}^{\prime}$ in a dense medium ({\it i.e.,} 
with $\pi_{T,L}\gg T^2,~m_{\gamma'}^2$) are explicitly given by
\begin{equation}
  e_{\rm eff}^e=\frac{(\varepsilon e+e')m_{\gamma'}^2}{\pi_{T,L}},
  ~
  e_{\rm eff}^p=-\frac{(\varepsilon e+e')m_{\gamma'}^2}{\pi_{T,L}},
  ~{\rm and}~
  e_{\rm eff}^n=e'.
\end{equation}
We observe that the dark vector couplings to the electrically charged fermions are suppressed by both the mixing term in the Lagrangian and medium-induced effects,
while the interaction of $A_{\mu}^{\prime}$ to the neutrons is not suppressed by the plasma effects.
As a result, there exists a suppression from the plasma effect in the emission of dark vectors from the electron and the proton currents.
On the other hand, the production of dark vectors via the neutron current does not have such a suppression,
and the emission rate from neutrons in the medium can be approximated by that in the vacuum.

For a NS of interest here, because of the large number of neutrons in a NS core, the dark vectors can be copiously produced 
via the bremsstrahlung process involving the neutron current.  Furthermore, since we are concerned with the case of low dark vector masses, 
the contributions from the electron and proton bremsstrahlung processes in the crust can and will be neglected.
Note that the nucleon superfluidity in the NS core is still under debate.  At temperatures below the critical temperature of Cooper pair 
formation, the nucleon bremsstrahlung rate may be suppressed by nucleon superfluidity. 
However, recent studies~\cite{Potekhin2020MNRAS} on NS cooling indicate that neutron superfluidity in the middle-aged NS core is weak 
and the critical temperatures are possibly too low to be relevant for our analyses. Thus, following Ref.~\cite{Buschmann2021PRL} the 
nucleon superfluidity effect will be ignored in our work.

% In the weak-mixing limit, Eq.~\eqref{eq:nonQED} can be solved by using the equivalent of time-dependent perturbation theory in quantum mechanics, 
% and lead to the following approximation for the dark vector-photon conversion probability
There exists oscillations between dark vector and photon due to the mixing term in our model. 
In the presence of an inhomogeneous external field, the approximation for the dark vector-photon conversion probability in the weak-mixing limit is given by
\begin{equation}
  \label{eq:probability}
  P_{\gamma' \rightarrow \gamma}=\varepsilon^{2}\left|\int_{r_0}^{r}dr^{\prime} \Delta\left(r^{\prime}\right) 
  \exp \left\{i \int_{r_0}^{r^{\prime}}dr^{\prime \prime}\left[\Delta\left(r^{\prime \prime}\right)-\Delta_{\gamma'}\right]\right\}\right|^{2}.
\end{equation}
The integral starts from the stellar surface $r_0=R_s$ since the x-ray photons produced inside the NS would be absorbed.
The parameters and detailed calculations for Eq.~\eqref{eq:probability} are presented in Appendix~\ref{apd:prob}.
For $m_{\gamma^{\prime}}\lesssim (T_s/R_s)^{1/2}$, we find an analytical formula of Eq.~\eqref{eq:probability}, which can be found in Appendix~\ref{apd:analyticalProb}.
We will see that the NS is an excellent test bed for the dark vector hypothesis since the conversion probability can be largely enhanced under a strong magnetic field.  
For the approximation to make sense, $m_{\gamma'}$ cannot take the vanishing limit and the numerical value for $\varepsilon$ should be sufficiently small to satisfy 
the weak-mixing condition.

A few remarks are in order.  The traditional $U(1)_{B-L}$ model is gauge-anomaly free if right-handed neutrinos are introduced.  
While anomaly free at higher energies, the effective Lagrangian~\eqref{eq:u1larg} is anomalous due to the mixing term at low energies.  
And it is the phenomenology of this low-energy effective Lagrangian~\eqref{eq:u1larg} that we want to discuss in this work.  We also note that the constraints 
on $e'$ that we obtain from NS cooling in this work do not depend on the mixing term and can be applied to the traditional $U(1)_{B-L}$ model.

%%%%%%%%%%%%%%%%%%%%%%%%%%%%%%%%%%%%%%%%%%%%%%%%%%
\section{Dark vector production}
\label{sec:dp}
%%%%%%%%%%%%%%%%%%%%%%%%%%%%%%%%%%%%%%%%%%%%%%%%%%

In this section, we calculate the emission rate of dark vectors from the neutron bremsstrahlung in the NS core. 
Supplements for the calculations are given in Appendix~\ref{apd:degenerate}.

%%%%%%%%%%%%%%%%%%%%%%%%%%%%%%%%%%%%%%%%%%%%%%%%%%
\subsection{Dark vector emission rate}
\label{eq:dper}
%%%%%%%%%%%%%%%%%%%%%%%%%%%%%%%%%%%%%%%%%%%%%%%%%%

The dark vector energy emission rate per unit volume for the process $N_1+N_2\to N_3+N_4+\gamma'$ ($N=n,~p$) in a strongly degenerate 
nuclear matter is given by~\cite{Brinkmann1988PRD,Harris2020JCAP}
\begin{equation}
  \label{eq:ee2}
  \begin{aligned}
  Q_{\gamma'} 
  =&
  \int \frac{d^{3} \mathbf{p}_{1}}{(2 \pi)^{3}} \frac{d^{3} \mathbf{p}_{2}}{(2 \pi)^{3}} 
  \frac{d^{3} \mathbf{p}_{3}}{(2 \pi)^{3}} 
  \frac{d^{3} \mathbf{p}_{4}}{(2 \pi)^{3}} \frac{d^{3} \omega}{(2 \pi)^{3}} \frac{S|\mathcal{M}|^{2}}{2^{5} 
  E_{1}^* E_{2}^* E_{3}^* E_{4}^* \omega} \omega f_{1} f_{2}\left(1-f_{3}\right)\left(1-f_{4}\right) \\
  & \times(2 \pi)^{4} \delta\left(E_{1}^*+E_{2}^*-E_{3}^*-E_{4}^*-\omega\right) \delta^{3}
  \left(\mathbf{p}_{1}+\mathbf{p}_{2}-\mathbf{p}_{3}-\mathbf{p}_{4}-\mathbf{p}_{\gamma'}\right),
\end{aligned}
\end{equation}
where the symmetry factor $S=1/4$ for the identical particles in the initial and final states and $S=1$ for mixed processes, 
the squared matrix element $|\mathcal{M}|^2=\sum_{\rm spins}|M|^2$ sums over initial and final spins, $\omega$ is the energy of 
emitted dark vector, $\mathbf{p}_k$ is the momentum associated with the nucleon $N_k$ ($k = 1, 2, 3, 4$), and $\mathbf{p}_{\gamma'}$ 
is the momentum of the dark vector.
In the presence of a nuclear mean field $U_N$, the energy dispersion relation of the nucleon is modified to be 
\begin{equation}
  \label{eq:enedisp}
  E_N=E^{*}+U_N=\sqrt{p^2+m_{*}^2}+U_N,
\end{equation}
where $m_{*}$ is the effective nuclear mass and $E^{*}= \sqrt{p^{2}+m_{*}^{2}}$ is the effective energy.
Note that the nuclear energies that enter the energy $\delta$-function and the denominator on the right-hand side of Eq.\eqref{eq:ee2} are the effective energies~\cite{Harris2020JCAP}.
The fermion phase-space distribution function is given by
\begin{equation}
  f_{k}=\left( e^{\left(E_k-\mu_k \right)/T}+1 \right )^{-1},
\end{equation}
where $\mu_k$ is the nuclear chemical potential of $N_k$.  With Eq.~\eqref{eq:enedisp} we have
$E_k-\mu_{k}=E_k^{*}-\mu_{k}^{*}$, where $\mu_{k}^{*}=\mu_{k}-U_N$.
The occupation numbers $f_{1,2}$ and the Pauli blocking factors $(1-f_{3,4})$ are for the incoming and outgoing nucleons. 
The dark vector are assumed to escape freely so that a Bose stimulation factor as well as its absorption are neglected.

In the nonrelativistic (NR) limit, the nucleon effective mass $m_*$ is much larger than all other energy scales such as the temperature or 
Fermi energies, so that the energy taken by the nucleon is $E^*\simeq m_*+E^{\rm kin}$, where $E^{\rm kin}=\mathbf{p}^2/2m_*$ is the 
nuclear kinetic energy. In a bremsstrahlung process, the radiation typically carries the momentum 
$|\mathbf{p}_{\gamma'}|\simeq E^{\rm kin}\ll |\mathbf{p}|$ and, therefore, the outgoing nucleons carry essentially all of the momentum and the radiation momentum can be neglected. 
With these approximations, we show in the next section the scattering matrix element of the nucleon bremsstrahlung, and in Appendix~\ref{apd:degenerate} we calculate the emission rate~\eqref{eq:ee2} in the degenerate limit.

%%%%%%%%%%%%%%%%%%%%%%%%%%%%%%%%%%%%%%%%%%%%%%%%%%
\subsection{Squared matrix element}
\label{sec:mat}
%%%%%%%%%%%%%%%%%%%%%%%%%%%%%%%%%%%%%%%%%%%%%%%%%%

As shown above, the effective couplings of $A_{\mu}^{\prime}$ to electron and proton in the medium are suppressed in the limit of light $m_{\gamma'}$.
The emission of dark vector from the neutron-neutron bremsstrahlung is the primary production mechanism in the inner core of a NS. 
The nucleon-nucleon interaction in the nucleon-nucleon bremsstrahlung was treated by the one-pion-exchange (OPE) approximation in Ref.~\cite{Friman1979APJ} 
and was used in the calculation of the axion emission rate from the nucleon-nucleon bremsstrahlung~\cite{Brinkmann1988PRD}.
For our present work, it is sufficiently reliable to calculate the dark vector emission rate using the OPE approximation (see also Ref.~\cite{Dent2012}).
Conservatively, we also include a factor of $1/4$~\cite{Beznogov2018PRC} in the emission rate to account for the overestimation in the OPE 
approximation at lower temperatures~\cite{Hanhart2001PLB}.
Additional remarks on the OPE approximation are provided in Appendix~\ref{apd:OPE}.

%*****************************fig1***************************************
\begin{figure}
%  \centerline{
  \includegraphics[width=100mm,angle=0]{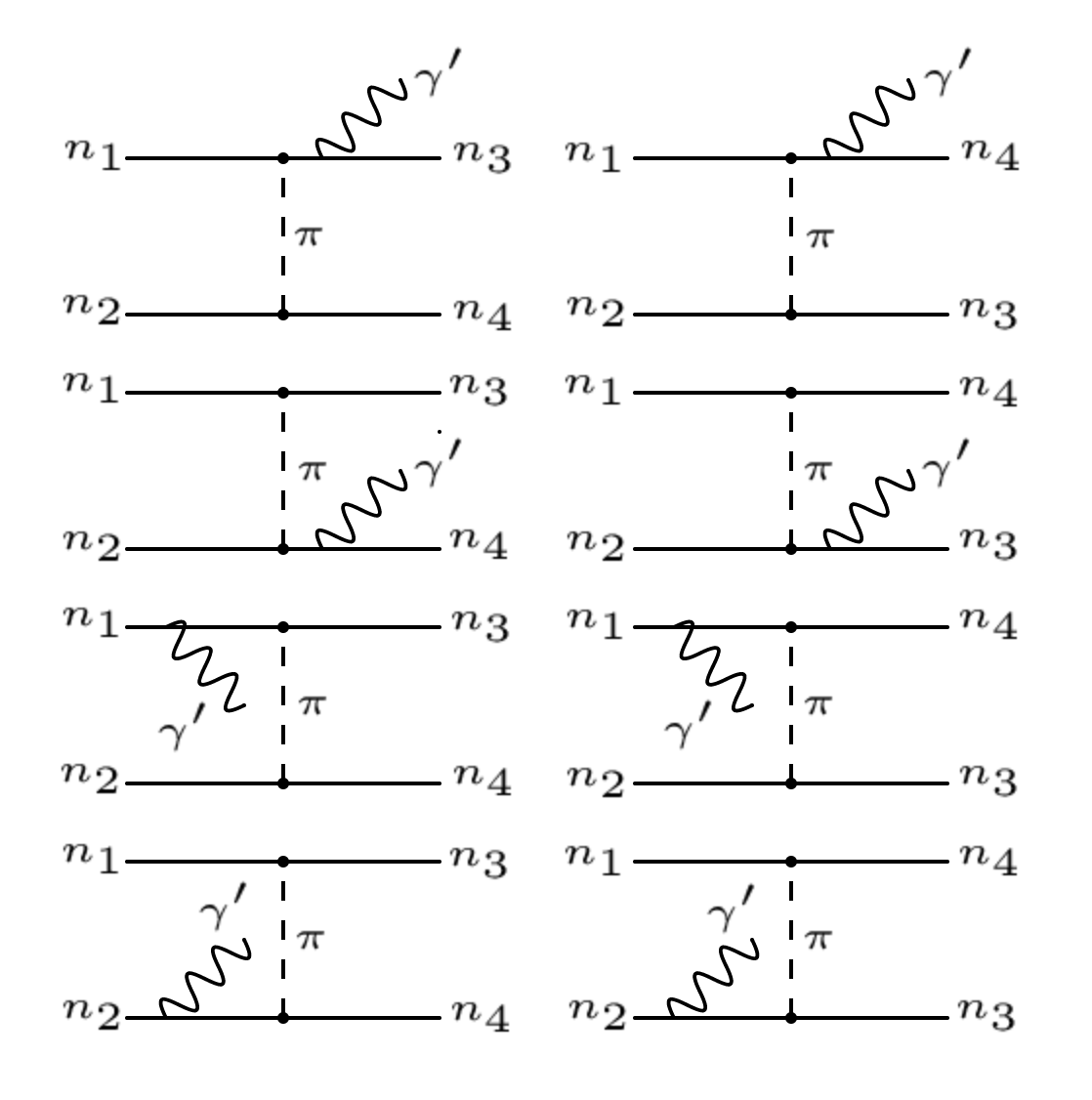}
  \caption{Feynman diagrams for the bremsstrahlung process $n+n\to n+n+\gamma'$.}
  \label{fig:fy1}
\end{figure}
%*************************************************************************

We show the Feynman diagrams for the process $n+n\to n+n+\gamma'$ (where $\gamma'\equiv A'$) in Fig.~\ref{fig:fy1}.
The dark radiation is induced by the neutron in the initial and final states in the scattering via the exchange of a neutral pion.  The squared matrix element is given by~\cite{Dent2012}
\begin{equation}
  \label{eq:SMpp}
  \begin{aligned}
  |\mathcal{M}|_{nn}^{2}&=\frac{64m_{*}^{2} |\mathbf{k}|^{2}}{\omega^{2}m_{\pi}^{4}}
  \left[\frac{C_{k}|\mathbf{k}|^{4}}{\left(|\mathbf{k}|^{2}+m_{\pi}^{2}\right)^2}+
  \frac{C_{l}|\mathbf{l}|^{4}}{\left(|\mathbf{l}|^{2}+m_{\pi}^{2}\right)^2}
  +\frac{C_{k l}\left(|\mathbf{k}|^{2}|\mathbf{l}|^{2}-2|\mathbf{k} \cdot \mathbf{l}|^{2}\right)}{\left(|\mathbf{k}|^{2}+
  m_{\pi}^{2}\right)\left(|\mathbf{l}|^{2}+m_{\pi}^{2}\right)}\right],
  \end{aligned}
\end{equation}
where $m_{\pi}$ is the pion mass, the momenta $k=P_2-P_4$, $l=P_2-P_3$, with $P_k$ being
the four-momenta associated with $n_k$ ($k = 2,3,4$), and $\mathbf{k}=\mathbf{p}_2-\mathbf{p}_4$ and $\mathbf{l}=\mathbf{p}_2-\mathbf{p}_3$.
In the NR limit, the nucleon mass is much larger than the temperature, the dark vector mass, and the transferred momenta, 
{\it i.e.,} $m_*^2\gg \mathbf{k}^2,~ \mathbf{l}^2$.  Furthermore, the four-momenta transferred among the nucleons are much larger 
than the momentum carried by the dark vector, {\it i.e.,} $k^2,~l^2\gg p_{\gamma'}^2,~\omega^2,~m_{\gamma'}^2$.
For the model considered in this work, the coefficients are given by
\begin{equation}
  C_{k}=f_{nn}^{4}\left(g_{\alpha}^{2}+g_{\beta}^{2}\right),~
  C_{l}=f_{nn}^{4}\left(g_{\alpha}^{2}+g_{\beta}^{2}-2 g_{\alpha} g_{\beta}\right),~
  C_{k l}=f_{nn}^{4}\left(g_{\alpha}^{2}+g_{\beta}^{2}-2 g_{\alpha} g_{\beta}\right),
\end{equation}
where $f_{nn} = f$ and $f\simeq 1.05$ is the pion-neutron coupling, 
the parameter $g_{\alpha}=g_{\beta}=e'\equiv g_n$ is the coupling between 
$A'$ and neutron. We see that only the coefficient $C_k$ is nonzero.

\begin{figure}
  %\centering
  \includegraphics[width=100mm,angle=0]{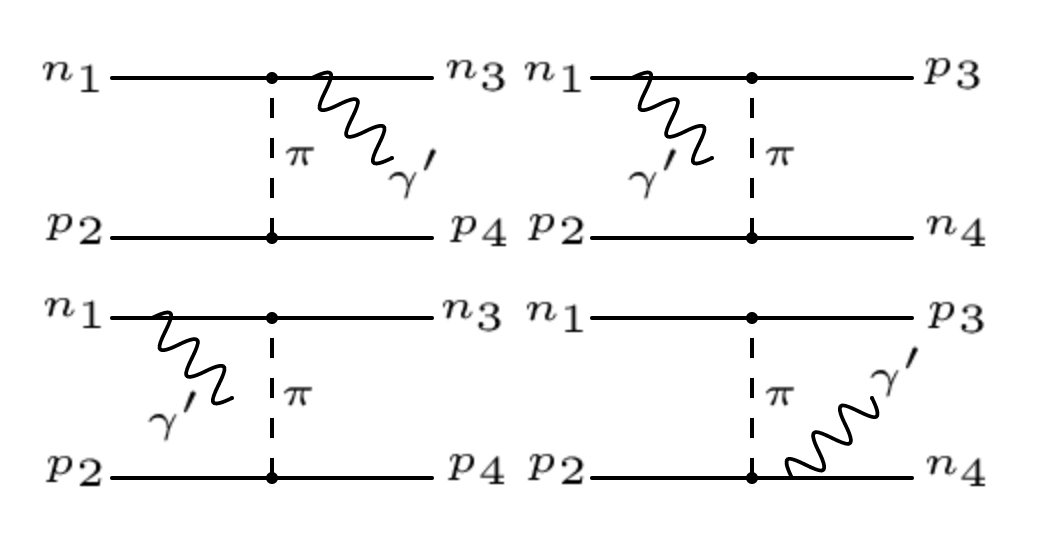}
  \caption{Feynman diagrams for the bremsstrahlung process $n+p\to n+p+\gamma'$.}
  \label{fig:fy2}
\end{figure}

The Feynman diagrams for the process $n+p\to n+p+\gamma'$ are depicted in Fig.~\ref{fig:fy2}. 
%There is no intermediate dark radiation diagrams since the pions have a zero baryon number so that they do not couple to the dark vector. 
The squared matrix element for this process is given by
\begin{equation}
  \label{eq:SMpn}
  |\mathcal{M}|_{np}^{2}=\frac{64 m_{*}^{2}|\mathbf{k}|^{2}}{\omega^{2} m_{\pi}^{4}}
  \left[\frac{C_{k}|\mathbf{k}|^{4}}{\left(|\mathbf{k}|^{2}+m_{\pi}^{2}\right)^{2}}+
  \frac{C_{l}|\mathbf{l}|^{4}}{\left(|\mathbf{l}|^{2}+m_{\pi}^{2}\right)^{2}}\right],
\end{equation}
where $C_k=C_l=f_{np}^4g_n^2$, with $f_{np}=\sqrt{2}f$ as required by isospin invariance. 
Due to the plasma effects, we have neglected the diagrams where the dark vector is emitted from an electron or proton.

%************************************************************************
%************************************************************************

%%%%%%%%%%%%%%%%%%%%%%%%%%%%%%%%%%%%%%%%%%%%%%%%%%
\subsection{Dark vector luminosity}
%%%%%%%%%%%%%%%%%%%%%%%%%%%%%%%%%%%%%%%%%%%%%%%%%%

% \subsubsection{Equation of State}

Given the emission rate, to compute the luminosity of dark vectors from the NS core we need to know the NS equation of state (EOS),
which determines the matter distribution in the NS, the chemical potential of the matter, and the profiles of neutron and proton Fermi momenta.
In our analysis, we employ the Akmal-Pandharipande-Ravenhall (APR) EOS~\cite{Akmal1998PRC} to model the uniform nuclear matter in the NS core 
and assume a NS mass of $1.4~M_{\odot}$, which corresponds to a NS core radius of $11.0$~km. 
It is expected that the nucleon bremsstrahlung in the NS core dominates the production of dark vectors.
For the weak coupling theory, the mean free path of dark vector in the medium is much larger than the size of the NS.  So in the calculation we 
have neglected the dark vector reabsorption effect. 
% We do not consider the possibility of exotic phases of matter in the core.

The differential dark vector luminosity is given by integrating the differential emissivity over the volume of the NS
\begin{equation}
  \label{eq:dldp}
  \frac{dL_{\gamma'}}{d\omega}=4\pi\int_{0}^{r_0}dr r^2\frac{dQ_{\gamma'}}{d\omega}.
\end{equation}
Integrating the differential luminosity over $\omega$, we obtain the total dark vector luminosity.  
With the APR EOS for the NS and a light dark vector with mass $m_{\gamma'}\ll $~keV, the dark vector luminosity from the NS core can be estimated as
\begin{equation}
  \label{eq:DVLumi}
  L_{\gamma'}\simeq \left(3.0\times 10^{37}~\mathrm{erg}\cdot\mathrm{s}^{-1}\right)
  \left( \frac{e'}{10^{-12}}\right)^{2}\left( \frac{T_c}{10^{9}\mathrm{~K}}\right)^{4},
\end{equation}
where $T_c$ is the core temperature of the NS.
The emission of dark vectors can result in NS cooling. Since the standard NS cooling scenario in terms of neutrino and photon emissions fits 
rather well to the observation of the NSs, the stellar cooling can be used to constrain the dark vector model.

%*****************************fig1***************************************
\begin{figure}
  %\centering
  \includegraphics[width=110mm,angle=0]{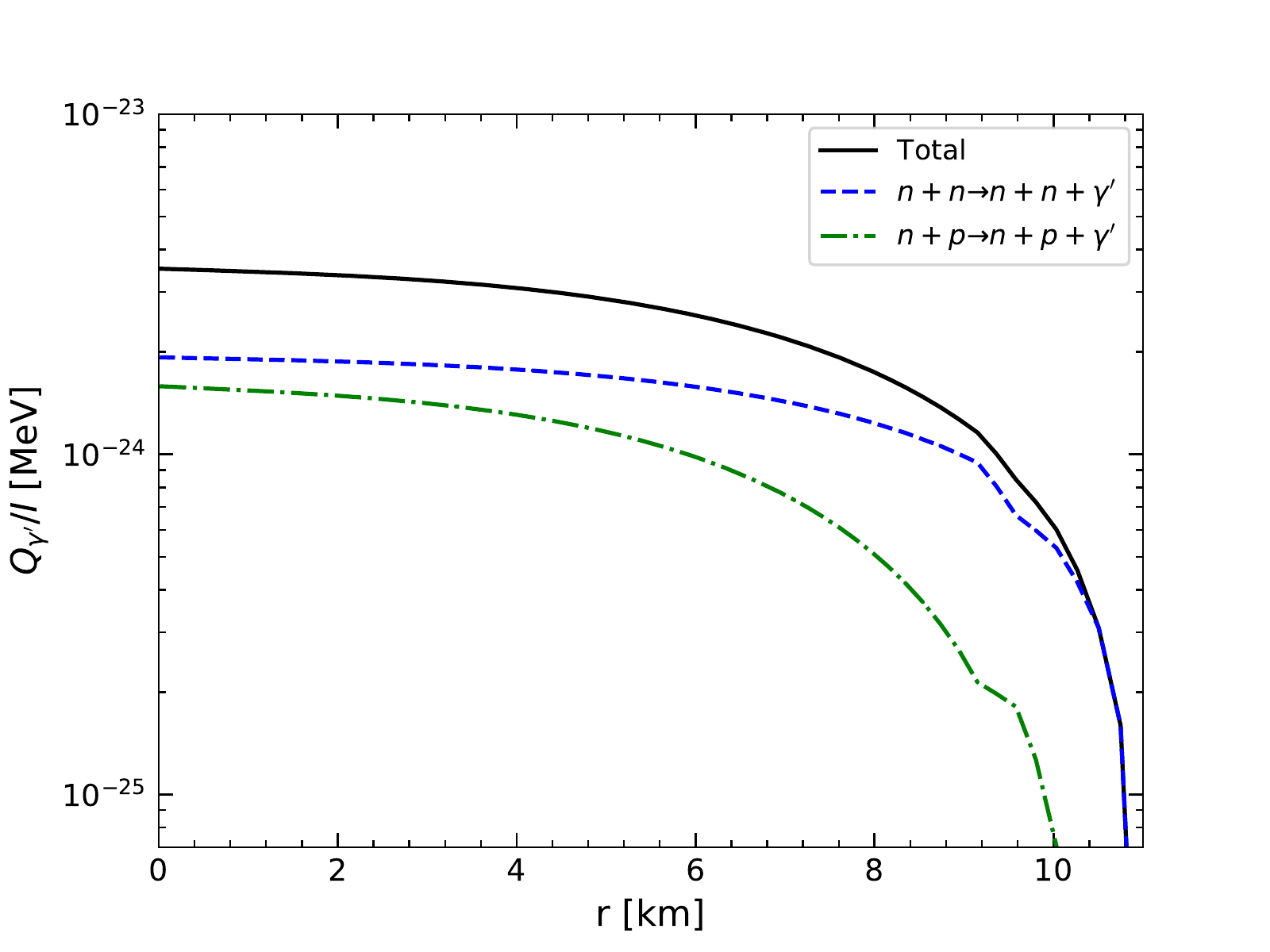}
  \caption{The $Q_{\gamma'}/I$ profile as a function of the radius in the core of the NS, with $e'=10^{-12}$ and $m_{\gamma'}=10^{-5}$~eV.}
  \label{fig:qprofile}
\end{figure}
%*************************************************************************

\begin{figure}
  %\centering
  \includegraphics[width=75mm,angle=0]{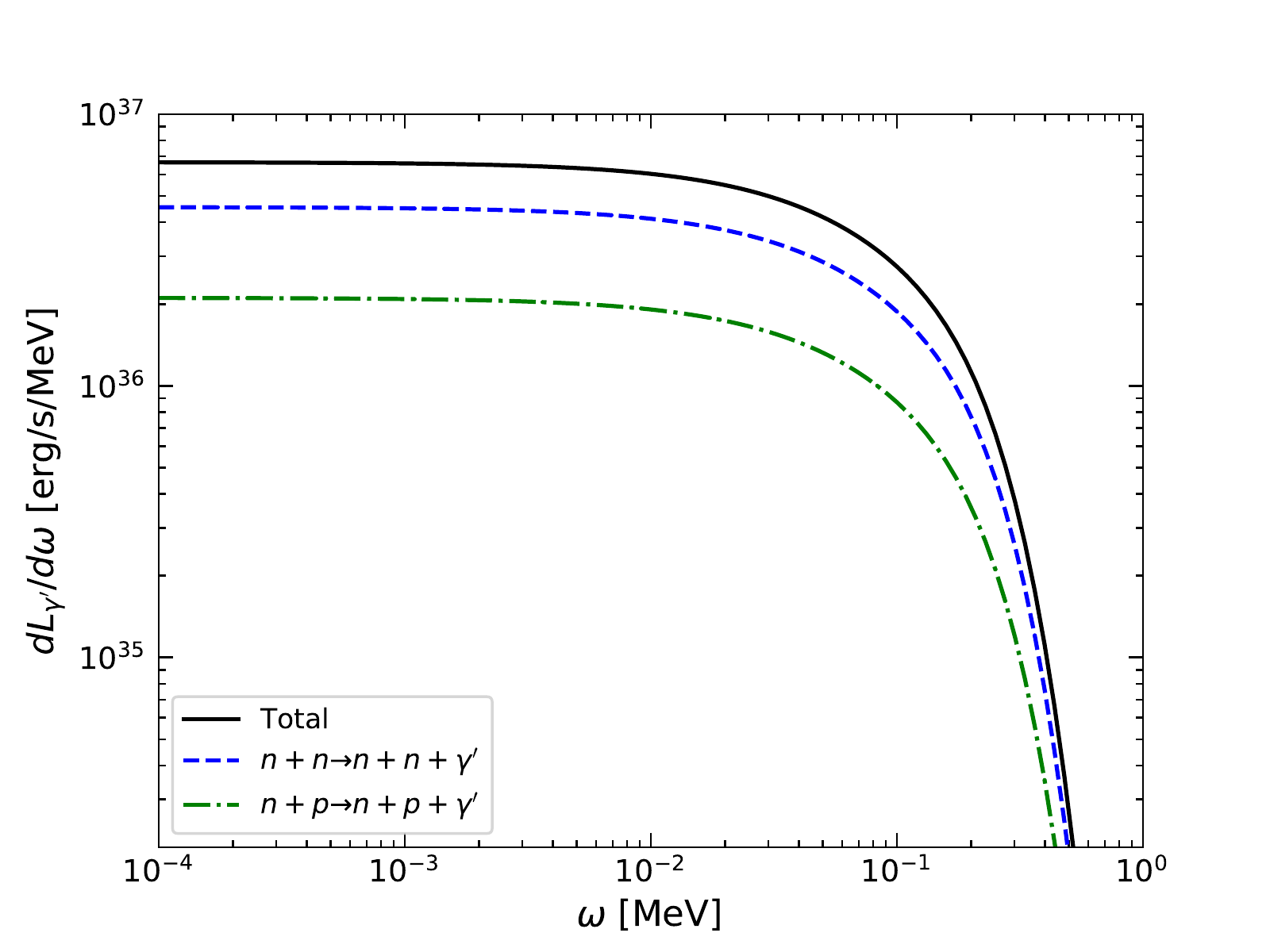}
  \includegraphics[width=75mm,angle=0]{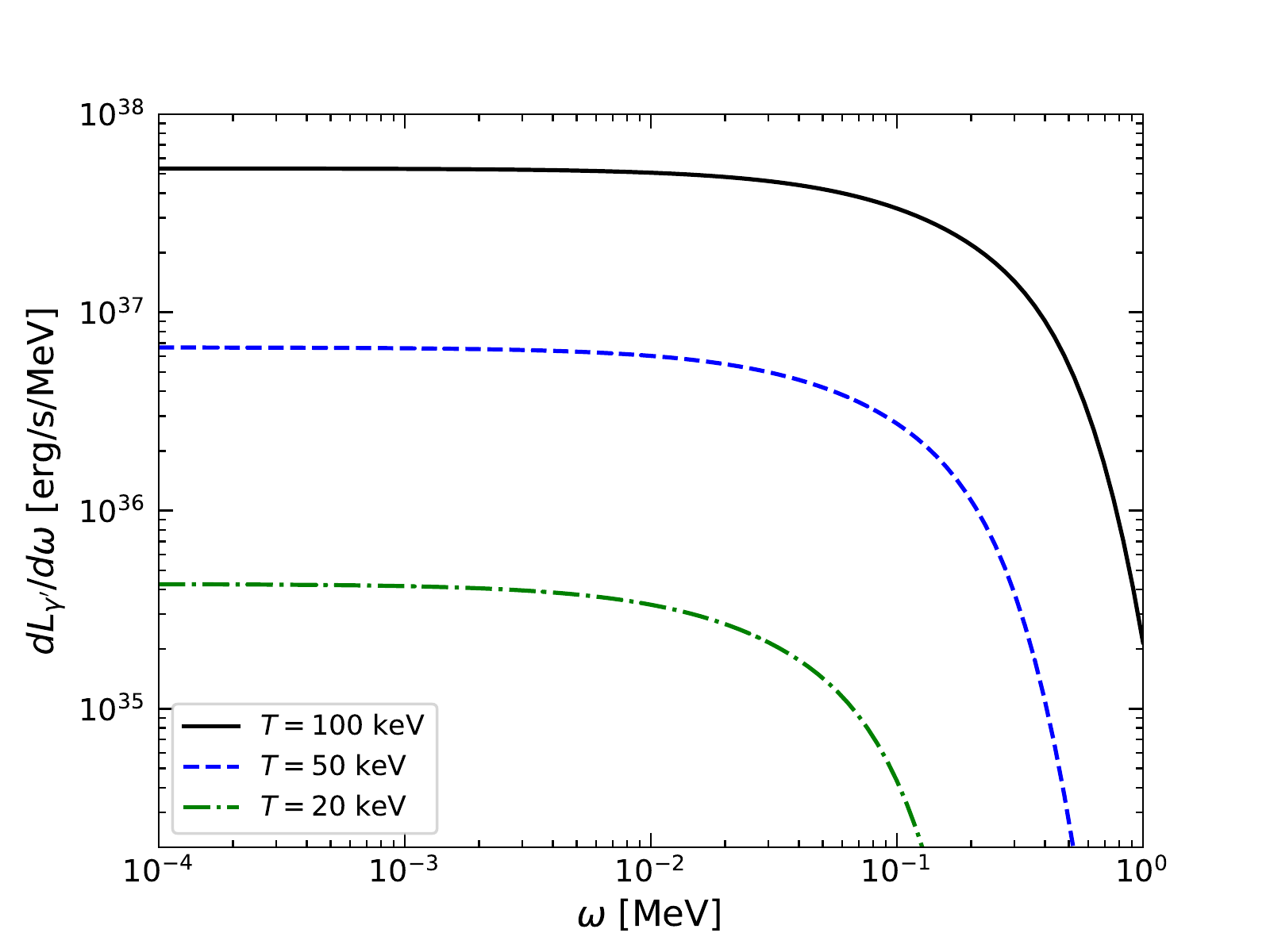}
  \caption{The dark vector spectrum $dL_{\gamma'}/d\omega$ as a function of frequency $\omega$, with $e'=10^{-12}$ and $m_{\gamma'}=10^{-5}$~eV.  We fix the core temperature at $T_c=50$~keV in the left plot.  The dot-dashed, dashed, and solid lines in the right plot represent the total
  spectrum with the core temperature $T_c=10,~50,~100$~keV.}
  \label{fig:dpdl}
\end{figure}

In Fig.~\ref{fig:qprofile}, we plot the dark vector ``energy emission rate'' $Q_{\gamma'}/I$, where $I$ is given by Eq.~\eqref{eq:eneint}, 
as a function of the NS radius, with $m_{\gamma'}=10^{-5}$~eV and $e'=10^{-12}$. The dashed and dot-dashed curves represent the results for the 
process $n+n\to n+n+\gamma'$ and $n+p\to n+p+\gamma'$, respectively. As expected, the neutron-neutron bremsstrahlung dominates the emission of 
dark vectors. The nearly constant emission rate extends to the radius $r\lesssim 8$~km and decreases sharply when approaching the surface.
We show the differential luminosity of dark vectors as a function of the dark vector energy in Fig.~\ref{fig:dpdl}, again with 
$m_{\gamma'}=10^{-5}$~eV and $e'=10^{-12}$. 
As shown in the left plot of Fig.~\ref{fig:dpdl}, the differential luminosity of dark vectors is nearly a constant 
for $\omega\lesssim T_c$ and cuts off at $\omega\sim T_c$ due to the factor $e^{-\omega/T_c}$ in the differential luminosity.
In the right plot of the figure, we observe that both the constant regime and the cutoff point of 
$dL_{\gamma'}/d\omega$ increase with the NS core temperature.

%*****************************fig1***************************************
%*************************************************************************

%*************************************************************************
\section{Neutrino and photon luminosities}\label{sec:nplum}
%*************************************************************************

One of the most fascinating stars known in the Universe is the NS, whose birth in supernova explosions is accompanied by the most powerful 
neutrino outburst.
The standard NS cooling scenario based upon the pioneering works~\cite{Chiu1964PRL,Bahcall1965PLBa,Bahcall1965PLBb,Friman1979APJ,Maxwell1987APJ} 
includes several neutrino emission processes (see~\cite{Yakovlev2001PR} for a review). 
The direct Urca cooling process consists of beta decay and electron capture that can induce a huge sink of energy in the NS core. 
However, a sufficiently small separation between the Fermi levels of protons and neutrons is required to activate this mode, which generally requires a minimum mass of the NS, denoted by $M_{\rm D}$, that is 
significantly larger than the canonical NS mass of $1.4~M_{\odot}$~\cite{Lattimer1991PRL,Beloborodov2016APJ}. 
For the NS with mass less than $M_{\rm D}$, the modified Urca cooling mode can take place everywhere and dominate the NS cooling process.
The modified Urca process is similar to the direct Urca, but involves an additional nucleon spectator,
\begin{eqnarray}
  \label{eq:UrcaNB}
  && n+n \rightarrow p+n+e+\bar{\nu}_{e}, \quad n+p+e \rightarrow n+n+\nu_{e},\\
  \label{eq:UrcaPB}
  && p+n \rightarrow p+p+e+\bar{\nu}_{e}, \quad p+p+e \rightarrow n+p+\nu_{e}.
\end{eqnarray}
The modified Urca by the neutron branch is shown by Eq.~\eqref{eq:UrcaNB}. Its neutrino emission rate in the NS core is given by~\cite{Friman1979APJ}
\begin{equation}
  Q_{\nu}^n\simeq\left(7 \times 10^{20}~\mathrm{erg}\cdot\mathrm{s}^{-1}\cdot\mathrm{cm}^{-3}\right)R_{\rm M}
  \left(\frac{\rho}{\rho_{0}}\right)^{\frac{2}{3}}\left(\frac{T_c}{10^{9}\mathrm{~K}}\right)^{8},
\end{equation}
where $\rho$ is the NS mass density profile, $\rho_{0}=2.8 \times 10^{14} \mathrm{~g} \cdot \mathrm{cm}^{-3}$ is the nuclear saturation density,
$T_c$ is the NS core temperature, and the suppression factor $R_{\rm M}\leq 1$ appears with the onset of superfluidity. 
Cooper pair cooling could become dominant if the superfluidity occurs. However, as argued above, the critical temperature for the Cooper pair 
formation is possibly too low to be relevant for this work.  We thus do not take the superfluidity into account.  

The proton branch, Eq.~\eqref{eq:UrcaPB}, was first analyzed in Ref.~\cite{Yakovlev1995AA} by using the same formalism as that in Ref~\cite{Friman1979APJ}.
The expression for the neutrino emissivity in the proton branch can be approximated by the following rescaling relation~\cite{Yakovlev2001PR} 
\begin{equation}
  Q_{\nu}^p\simeq \left(\frac{m_{*}^p}{m_*^{n}}\right)^{2} \frac{\left(p_{{F,e}}+3 p_{{F},p}-p_{{F,n}}\right)^{2}}{8 p_{{F,e}} p_{{F},p}} \Theta_{F}Q_{\nu}^n,
\end{equation}
where $m_*^p$ and $m_*^n$ are the effective masses of proton and neutron, respectively, and $p_{F,e}$, $p_{F,p}$, and $p_{F,n}$ are the Fermi momenta of 
electron, proton, and neutron, respectively. Note that $\Theta_{F}=1$ for $p_{{F},n}<3 p_{{F},p}+p_{{F},e}$; otherwise, $\Theta_{F}=0$.
As an example, take $p_{{F,e}}=p_{{F},p}=p_{{F,n}}/2$, we have $Q_{\nu}^p\simeq 0.5 Q_{\nu}^n$, 
i.e., the proton branch is nearly as efficient as the neutron branch. In addition to the Urca processes, the neutrino's emission processes can also be 
induced by the bremsstrahlung in nucleon-nucleon collisions, which, however, are subdominant for the NS cooling (a summary of these processes can be
found in Figs 11 and 12 of Ref.~\cite{Yakovlev2001PR}). Summing up all of the processes, the total neutrino emission rate can be estimated as 
$Q_{\nu}\simeq RQ_{\nu}^n$, with a rescaling factor $R\simeq 1.5$~\cite{Yakovlev2001PR}.
With the neutrino emission rate, we can then determine the neutrino luminosity by the integration 
\begin{equation}
  L_{\nu}=4\pi\int_0^{r_0}drr^2Q_{\nu},
\end{equation}
where $r_0=11$~km is the NS core radius. For the APR EOS, the neutrino luminosity is then given by 
\begin{equation}
  \label{eq:NuLumi}
  L_{\nu}\simeq \left(8.1\times 10^{39}~\mathrm{erg}\cdot\mathrm{s}^{-1}\right)\left(\frac{T_c}{10^{9}\mathrm{~K}}\right)^{8}.
\end{equation}

\begin{figure}
  %\centering
  \includegraphics[width=110mm,angle=0]{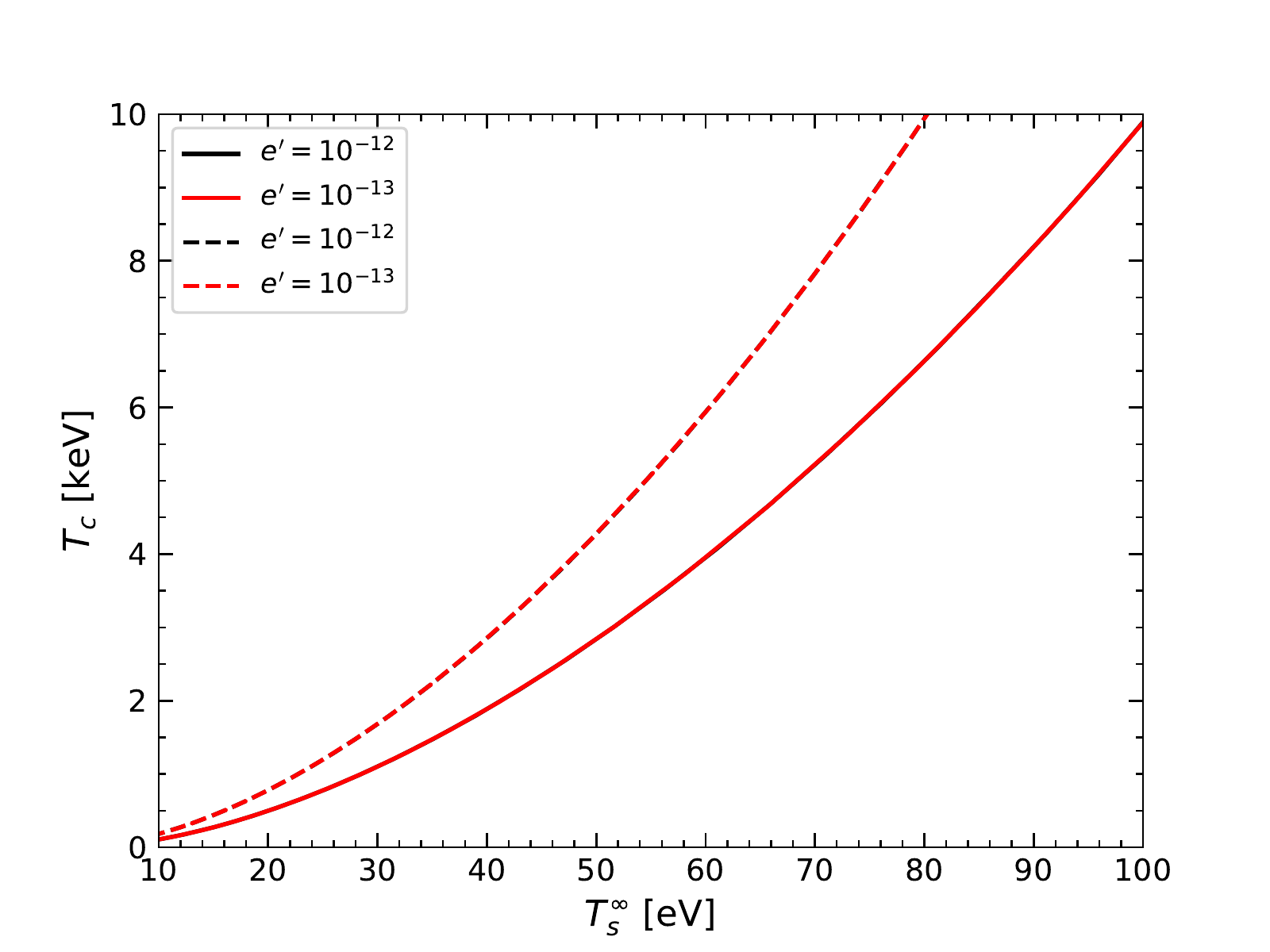}
  \caption{We determine the relation between the core temperature and the surface temperature at infinity by the {\tt NSCool} code. 
  The black and red lines represent the results with $e^{\prime}=10^{-12}$ and $10^{-13}$, respectively, and coincide with each other in the figure. 
  The dashed and solid lines denote the results with and without taking into account the pair breaking formation (PBF) 
  processes, respectively. We take the $\gamma-\gamma^{\prime}$ mixing angle $\varepsilon=0$ in this plot.}
  \label{fig:TeTc}
\end{figure}

Previous literature has obtained the constraints on new physics models, e.g., the axion coupling, by requiring that the emissivity of the new particle should not exceed that of the neutrino, i.e., $L_{\gamma^{\prime}}<L_{\nu}$~\cite{Raffelt1996}. 
Otherwise, the evolution of the NS will be strongly altered by the new physics, which is in conflict with the observations.
As shown above, in addition to the NS density profile, the luminosities of dark vector and neutrino strongly depend on the core temperature,
which, however cannot be determined by the EOS or by direct measurements.
The core temperature may be inferred from the NS surface temperature observations but suffers from large uncertainties~\cite{Buschmann2021PRL}.
In Fig.~\ref{fig:TeTc}, we depict the relation between core temperature and surface temperature at infinity 
using {\tt NSCool} code~\cite{Page2016} (with the default NS model).
We observe that the relation is independent of the new gauge coupling $e^{\prime}$ (with $\varepsilon=0$). However, as we will show below,
the luminosity at infinity cannot be solely determined by the luminosity at NS surface if there are conversions between photon and dark vector during
their propagation toward the observer. In this case, there is no direct relation between core temperature and surface temperature at infinity.
Furthermore, in the standard NS evolution, the additional heating effects can be neglected for the middle-aged NS, and the evolution of the NS core 
temperature with the time is expected to follow a power law $T_c\simeq (10^{9}\mathrm{~K})\left(t/\mathrm{yr}\right)^{-1/6}$~\cite{Beloborodov2016APJ}.
Such a relation may also break down in new physics if the cooling of the NS is strongly affected by the emissions of new weakly-interacting particles.
In this work, we will perform numerical simulations of the NS cooling based on the modified {\tt NSCool} code that includes additional energy loss via the dark vector emission. In this way, we determine the core temperature and the surface luminosity for the NS with a given age.
This will be described in detail below.

%*****************************fig1***************************************
%*************************************************************************

In the standard NS cooling model, the emissions of neutrinos and photons lead to the NS cooling. 
In the early stage, the neutrino emission is the dominant mode for the NS cooling.
As the stellar temperature drops, the NS cooling by photon emissions becomes significant since the neutrino luminosity 
decreases much faster than that of photon.
For NSs with age $t\gtrsim 100$~kyr, the photon emission will exceed the neutrino one~\cite{Raffelt1996}, 
and thus dominates the NS cooling process.

We now turn to the photon emission of the NS. The NS cooling due to the emission of photons is directly measured by 
the NS surface photon luminosity, which is given by
\begin{equation}\label{eq:SBlaw}
  L_{\gamma}=4\pi R_{s}^2\sigma T_s^4,
\end{equation}
where $\sigma$ is the Stefan-Boltzmann constant, $R_s$ is stellar radius, and $T_s$ is stellar surface temperature.
Without photon-dark vector conversions, the observed surface photon luminosity at infinity is accordingly redshifted as
\begin{equation}
  L_{\gamma}^{\infty}=e^{2 \phi_{s}}L_{\gamma},
\end{equation}
where ${\phi_{s}}$ is the gravitational potential at the stellar surface, and
\begin{equation}
  e^{\phi_{s}}=\left(1-\frac{2GM}{R_s c^{2}}\right)^{1/2}\simeq 0.79,
\end{equation}
where $G$ is the gravitational constant, and $M$ is the stellar mass. 

However, the observed surface photon luminosity will be changed if the photons and dark vectors can convert into each other during their propagation to the Earth. For the NS with age $\sim 10^6$~yr that is concerned in this work, we will show that the surface luminosity of the NS will be 
dominated by the photon emissions. 
For the parameter space we are interested in, the dark vectors' luminosity is always a subdominant component and, therefore, their contributions to
the observed photon luminosity will be neglected. In this case, the surface photon luminosity observed at infinity is given by
\begin{equation}
  L_{\gamma}^{\infty}=e^{2 \phi_{s}}\int_{0}^{\infty}d\omega\left( 1-P_{\gamma\to\gamma^{\prime}}(\omega)\right)\frac{dL_{\gamma}}{d\omega},
\end{equation}
where $P_{\gamma\to\gamma^{\prime}}(\omega)=P_{\gamma^{\prime}\to\gamma}(\omega)$ is the photon-dark vector conversion probability, which will be 
given in Appendix~\ref{apd:prob}. Since the Stefan-Boltzmann law~\eqref{eq:SBlaw} can be obtained by integrating the Planck's law over $\omega$,
the differential surface luminosity of the photon is given by
\begin{equation}
  \frac{dL_{\gamma}}{d\omega}=4\pi R_s^2\frac{T_s^3}{4\pi^2}\frac{x^3}{e^x-1},~~{\rm with}~~x=\frac{\omega}{T_s}.
\end{equation}
We will show in Appendix~\ref{apd:analyticalProb} that the $\gamma-\gamma^{\prime}$ conversion probability can be written as $P_{\gamma\to\gamma^{\prime}}(\omega)=P_0\omega^2$ 
for dark vector in the mass range $m_{\gamma^{\prime}}\lesssim (\omega/R_s)^{1/2}\sim 3\times 10^{-5}$~eV, with $\omega\sim T_s\sim 50$~eV.
With these, we finally obtain the observed surface photon luminosity at infinity by taking into account the $\gamma-\gamma^{\prime}$ conversion
\begin{equation}\label{eq:lumsinf}
  L_{\gamma}^{\infty}=e^{2 \phi_{s}}\left( L_{\gamma}-\frac{2\pi^4P_0A_s}{63}T_s^6 \right),
\end{equation}
where $A_s=4\pi R_s^2$. The surface temperature at infinity is then given by 
\begin{equation}\label{eq:Tsinf}
  T_{s}^{\infty}=e^{ \phi_{s}}\left( T_s^4-\frac{2\pi^4P_0}{63}T_s^6 \right)^{1/4}.
\end{equation}
Using these results, we can determine the observed surface luminosity and temperature by including the redshift by the gravitational potential as well as the $\gamma-\gamma^{\prime}$ conversion once we know the values of $L_{\gamma}$ and $T_s$ at the NS surface.

The validation of Eqs.~\eqref{eq:lumsinf} and \eqref{eq:Tsinf} should be further clarified.
One may be concerned that the approximation of the probability requires $\omega\gtrsim m_{\gamma^{\prime}}^{2}R_s$.  However,
we have integrated over $\omega$ in the range of $(0,\infty)$ for $P_{\gamma\to\gamma^{\prime}}dL_{\gamma}/d\omega$.
It is easy for the reader to confirm that there is nearly no difference between the integration over $\omega$ in the ranges of $(0,\infty)$ and $(0.9T_s,\infty)$,
which means that the small values of $\omega\lesssim 0.9T_s$ make negligible contributions to the integration.
We thus conclude that Eqs.~\eqref{eq:lumsinf} and \eqref{eq:Tsinf} are valid for our calculations provided that 
$m_{\gamma^{\prime}}\lesssim 3\times 10^{-5}$, with $\omega\sim T_s\sim 50$~eV.

\begin{figure}
  %\centering
  \includegraphics[width=110mm,angle=0]{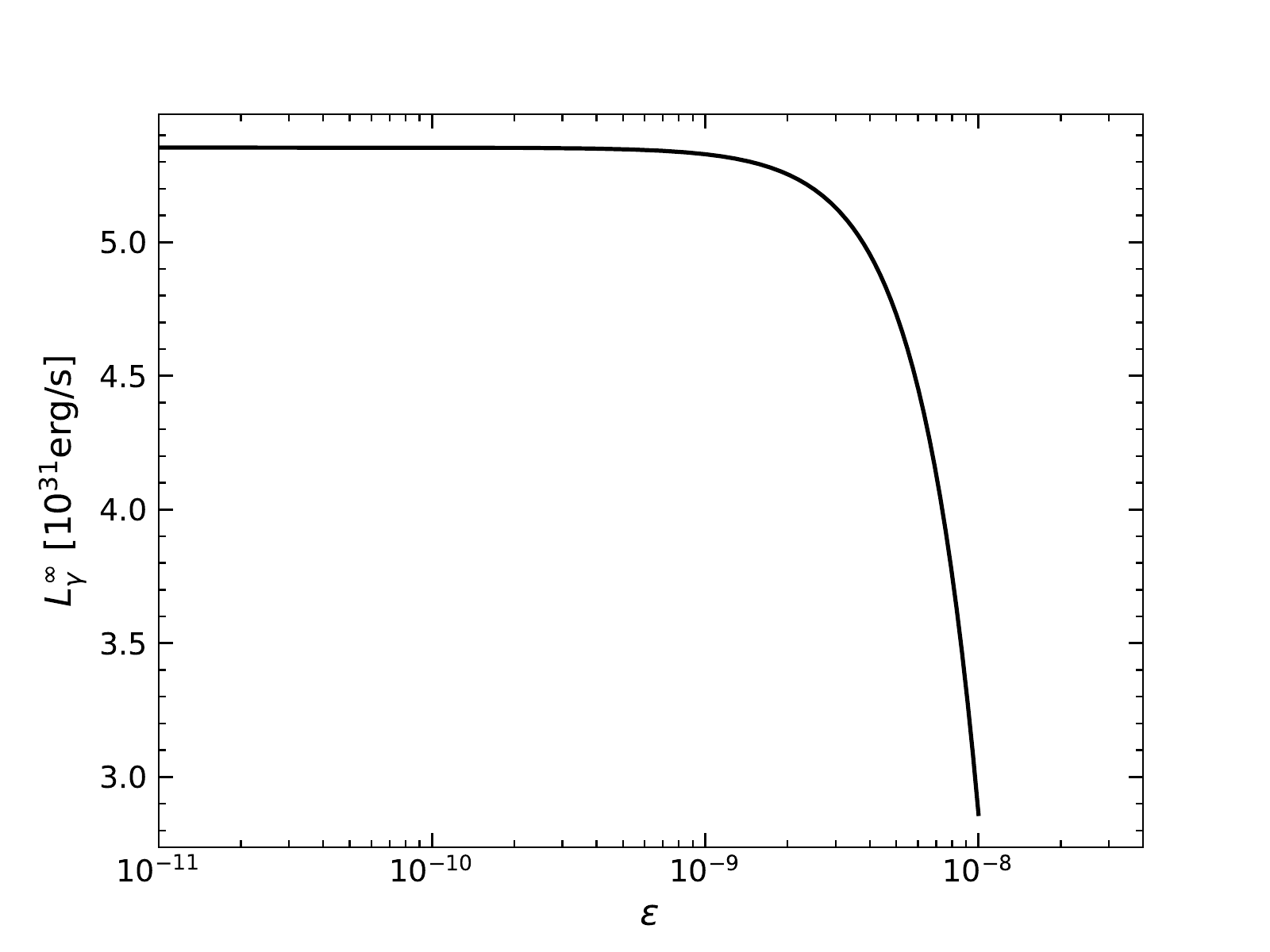}
  \caption{ Surface luminosity observed at infinity as a function of $\varepsilon$, with $L_{\gamma}=8.4\times 10^{31}$~erg/s and $T_s=47.1$~eV.}
  \label{fig:epsLum}
\end{figure}

In Fig.~\ref{fig:epsLum}, we show the variation of the surface luminosity observed at infinity $L_{\gamma}^{\infty}$ with the mixing 
angle $\varepsilon$. We assume the surface luminosity $L_{\gamma}=8.4\times 10^{31}$~erg/s and temperature $T_s=47.1$~eV for the NS.
As shown in this figure, the $\gamma-\gamma^{\prime}$ conversion strongly reduces the observed surface luminosity when the mixing angle
reaches values of $\sim 10^{-8}$.

%*****************************fig1***************************************
%*************************************************************************

%%%%%%%%%%%%%%%%%%%%%%%%%%%%%%%%%%%%%%%%%%%%%%%%%%
\section{NS cooling simulation}\label{sec:coolSimulate}
%%%%%%%%%%%%%%%%%%%%%%%%%%%%%%%%%%%%%%%%%%%%%%%%%%

In this work, we employ the {\tt NSCool} code~\cite{Page2016} for the numerical simulation of the thermal evolution of the star.
% solution of the Tolman-Oppenheimer-Volkoff (TOV) equation as well as
The current version of the {\tt NSCool} code incorporates all the corresponding neutrino cooling reactions, including direct and modified Urca processes,
nucleon-nucleon bremsstrahlung, as well as the thermal Cooper pair breaking and formation (PBF).
We modify the code to include the emissivity of the dark vector so that it can take part in the NS cooling along with the neutrino emission processes.
For the core of the NS, we adopt the APR EOS with a mass $1.4$~$M_{\odot}$ 
({\tt Prof\_APR\_Cat\_1.4.dat}). For the crust EOS, we use the default profile implemented by the file {\tt Crust\_EOS\_Cat\_HZD-NV.dat}.
% It has been pointed out that the public version of the code does not includes the collective correction due to anomalous terms 
% in the axial channel, which additionally significantly reduces the PBF neutrino emissivity. It was shown that the collective correction
% is very important for explaining the observed rate of change in the Cas A NS temperature. 

We focus on the observations from one of the M7 members, RX J1856.5-3754 (J1856 for short), which locates at a distance $123\pm 13$~pc away from us 
and has an age around $(4.2\pm 0.8)\times 10^5$~yr~\cite{Kerkwijk2008PAJ,Walter2010APJ} (a summary can be found in Refs.~\cite{webNScool,Potekhin2020MNRAS}).
% \footnote{A summary of the observed NSs' information can be found on the website~\url{http://www.ioffe.ru/astro/NSG/thermal/cooldat.html}.}.
Recent studies~\cite{Potekhin2020MNRAS} on NS cooling indicate that the critical temperature is very low and the neutron superfluidity is 
very weak for the middle-aged NS. We thus turn off the PBF processes in the simulations with {\tt NSCool}, as is done in Ref.~\cite{Buschmann2021PRL}.
We find that the results from the standard NS cooling without including the PBF processes can account for the surface luminosity observations 
on J1856 quite well. However, the simulation including the PBF processes predicts a much lower surface luminosity than the observation.

%*****************************fig1***************************************
\begin{figure}
  %\centering
  \includegraphics[width=110mm,angle=0]{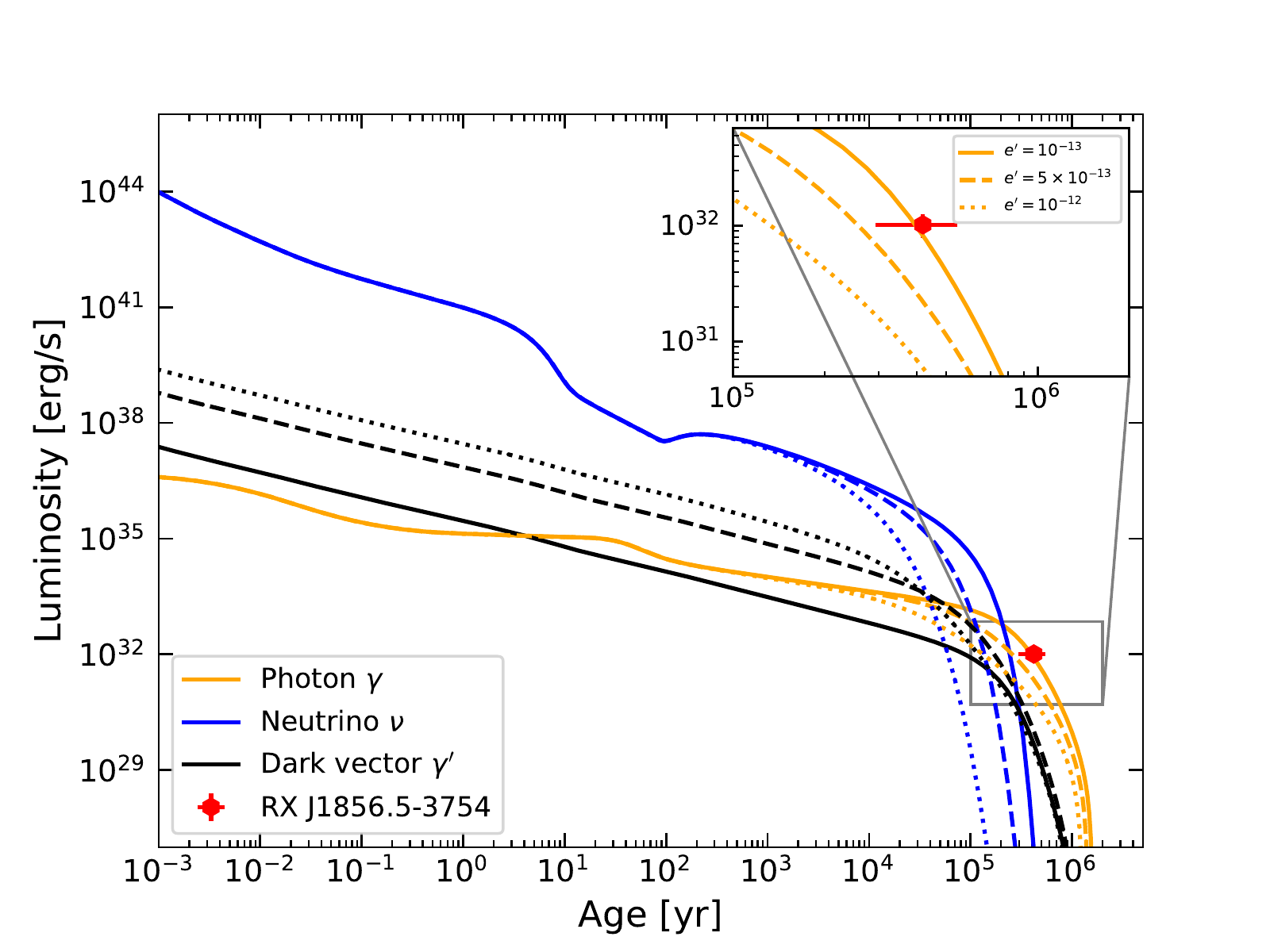}
  \caption{The NS surface luminosities from the simulations with $e^{\prime}=10^{-13}$ (solid line), $5\times 10^{-13}$ (dashed line), 
  and $10^{-12}$ (dotted line). The blue, dark, and yellow lines represent the neutrino, dark vector, and photon surface luminosities, respectively.
  The red point denotes the value of $L_{s}^{\infty}/e^{2\phi_{s}}$, where $L_{s}^{\infty}$ is the observed surface luminosity of J1856 at 
  infinity, and $e^{2\phi_{s}}$ is the redshift factor.}
  \label{fig:lumComponent}
\end{figure}
%*************************************************************************

In Fig.~\ref{fig:lumComponent}, we depict the neutrino (blue), dark vector (dark), as well as photon (yellow) surface luminosities as a 
function of the NS age. In the simulations, we assume $e^{\prime}=10^{-13},~5\times 10^{-13}$, and $10^{-12}$, and the results are represented 
by the solid, dashed, and dotted curves, respectively. The blue, dark, and yellow curves represent the neutrino, dark vector, and photon surface luminosities, respectively.
The red point represents the observation of the J1856 surface luminosity. 
It is shown by the simulations that the photon luminosity is the dominant component for J1856.
% for NS age $\lesssim 3\times 10^{5}$~yr, the neutrino luminosity dominates the cooling of the NS. On the other hand, for the older NS the photon emission
% plays the most significant role. As such, the photon luminosity is the dominate component for J1856.
We find that the model with a new gauge coupling $e^{\prime}\lesssim 10^{-13}$ predicts the same age-luminosity profile as the standard NS cooling model 
since in this case the cooling by the dark vector emission is negligible.  There is a significant deviation in the photon luminosity from the observation when $e^{\prime}$ increases to $5\times 10^{-13}$ and, therefore, $e^{\prime}$ should be constrained by the observation. We will show the constraints on the model by fitting to 
the observations below.

%%%%%%%%%%%%%%%%%%%%%%%%%%%%%%%%%%%%%%%%%%%%%%%%%%
\section{x-ray from dark vector-photon conversion}
\label{sec:plum}
%%%%%%%%%%%%%%%%%%%%%%%%%%%%%%%%%%%%%%%%%%%%%%%%%%

Recent analyses of the data from the XMM-Newton and Chandra x-ray telescopes~\cite{Dessert2020APJ} show a significant excess of hard x-ray 
emissions, in the energy range of $2-8$~keV, from the nearby M7 x-ray dim isolated NSs.  
In particular, it has been shown in Ref.~\cite{Dessert2020APJ} that the NS J1856 hard x-ray spectrum has a $\sim 5\sigma$ excess, which is the 
most significant hard x-ray excess among the M7 members.
The fact that the hard x-ray excess is observed in some NSs but not others depends on the experimental measurements as well as the NS properties~\cite{Buschmann2021PRL}.
Observations by the ROSAT All Sky Survey~\cite{Haberl2007ASS} have shown that all the M7 members have soft spectra that are well described by near-thermal 
distributions with surface temperatures $\sim 50-100$~eV and, therefore, they are expected to produce negligible hard x-ray flux. 
The explanation of the hard x-ray excess in the context of an axion model has been explored in Ref.~\cite{Buschmann2021PRL} and is found to be consistent 
with the current constraints. In this work, we explore the interpretation for the excess in the dark vector scenario.

The conversion probability between dark vector and photon is proportional to the square of the magnetic field strength. Observations show that all the M7 NSs 
have strong magnetic fields with a characteristic dipole magnetic field strength around $\sim 10^{13}$~G.
Therefore the NSs provide an excellent place to test the $\gamma^{\prime}-\gamma$ oscillations.
Having calculated both the differential luminosity of dark vectors~\eqref{eq:dldp} and the dark vector-to-photon conversion 
probability $P_{\gamma'\to\gamma}$ in Appendix~\ref{apd:prob}, the differential flux of hard x-ray photons produced from dark vector-photon 
conversion is given by 
\begin{equation}
  F_{\gamma}(\omega)=\frac{P_{\gamma'\to\gamma}}{4\pi d^2}\frac{dL_{\gamma'}}{d\omega},
\end{equation}
where $d$ is the distance between the NS and Earth. The probability depends on the dipole magnetic field orientation $\theta$. 
In this work we adopt the $\theta$-averaged conversion probability $\bar{P}_{\gamma'\to\gamma}=\frac{1}{2\pi}\int_0^{2\pi}d\theta P_{\gamma'\to\gamma}(\theta)$ as an approximation 
for our calculations. In the limit of small dark vector mass, $m_{\gamma'}\lesssim (\omega/R_s)^{1/2}\sim 10^{-4}$~eV for energy $\omega\sim 1$~keV and NS 
radius $R_s\simeq 11$~km, $\bar{P}_{\gamma'\to\gamma}$ becomes independent of $m_{\gamma'}$ (see Eq.~\eqref{eq:prob4} in Appendix~\ref{apd:analyticalProb}).
Note that in order to compare with the observed hard x-ray spectrum, the differential flux $F_{\gamma}(\omega)$ is calculated at energy $\omega=E_{\rm obs}/e^{\phi_s}$, where 
$E_{\rm obs}$ is the observed photon energy. Then the differential flux observed at infinity is given by $F_{\gamma}^{\infty}(E_{\rm obs})=e^{3\phi_s}F_{\gamma}(\omega)$.
Compared with the photon differential luminosity $dL_{\gamma}/d\omega$ whose maximum value is obtained at energy $\omega\simeq 2.82T_s\sim 141$~eV, 
with a surface temperature $T_s\sim 50$~eV, the hard x-ray spectrum $F_{\gamma}(\omega)$ from $\gamma^{\prime}\to \gamma$ conversion has its maximum value at a much 
higher energy $\omega\simeq 3.31T_c\sim 6.6$~keV, with a core temperature $T_s\sim 2$~keV (the surface-core temperature relation can be found in Fig.~\ref{fig:TeTc}).
We thus expect the flux from $\gamma^{\prime}-\gamma$ conversion can make contributions to the hard x-ray excess in the energy range $2-8$~keV and at the same time
reproduce the correct spectrum shape.

%%%%%%%%%%%%%%%%%%%%%%%%%%%%%%%%%%%%%%%%%%%%%%%%%%
\section{Statistical analysis}
\label{sec:dataAnalysis}
%%%%%%%%%%%%%%%%%%%%%%%%%%%%%%%%%%%%%%%%%%%%%%%%%%

Our starting point for the statistical analysis of the J1856 data in the context of the dark vector model is the Bayesian inference~\cite{Bayesian1992}.
For the model consisting of a set of parameters $\vec{\theta}=\left\{\theta^{1}, \theta^{2}, \cdots, \theta^{n}\right\}$, 
the Bayes theorem is given by
\begin{equation}
  \mathcal{P}(\vec{\theta}\mid {\rm data })=\frac{\mathcal{P}({\rm data } 
  \mid \vec{\theta}) \cdot \mathcal{P}(\vec{\theta})}{\mathcal{P}(\rm{ data })},
\end{equation}
where $\mathcal{P}(\rm{ data })$ is the data probability, $\mathcal{P}(\vec{\theta})$ is the prior probability that indicates the degree of belief one
has before observing the data, and $\mathcal{P}(\vec{\theta}\mid {\rm data })$ is the conditional probability-density function that is 
a posterior probability representing the change in the degree of belief one can have after giving the measurement data. Note that
$\mathcal{P}({\rm data}\mid \vec{\theta})=\mathcal{L}(\vec{\theta})$ links the posterior probability to the likelihood of the data, 
where the likelihood function $\mathcal{L}(\vec{\theta})$ takes the form of
\begin{equation}
  \mathcal{L}(\vec{\theta})=\exp \left(-\chi^{2}(\vec{\theta})/2\right),
\end{equation}
where 
\begin{equation}
  \chi^{2}(\vec{\theta})=\sum_{i}^{m} \frac{\left(\lambda_{i}^{\exp }-
  \lambda_{i}^{\text {the }}\right)^{2}}{\sigma_{i}^{2}},
\end{equation}
where $\lambda_{i}^{\exp }$ denotes the experimental value with an uncertainty in the measurement $\sigma_{i}$, $m$ represents the number of data points,
and $\lambda_{i}^{\text {the }}$ denotes the predicted value for a given model.

For the dataset, we take into account the J1856 x-ray spectrum observations in $2-8$~keV~\cite{Dessert2020APJ}, $T_s^{\infty}$~\cite{Buschmann2021PRL},
$L_{\gamma}^{\infty}$~\cite{Ho2007MNRAS,Mignani2013MNRAS,Yoneyama2017PASJ}, and the distance measurements~\cite{Kerkwijk2008PAJ,Walter2010APJ}.
Noting that $L_{\gamma}^{\infty}$ inferred from the observations is in a narrow band $(0.05-0.08)\times 10^{33}$~erg/s, we take 
$L_{\gamma}^{\infty}=0.065\times 10^{33}$~erg/s as the central value and assume a Gaussian distribution for the luminosity. 
The parameter set is $\vec{\theta}=\{d,e^{\prime},\varepsilon \}$, where the distance $d$ enters the spectrum function. 
We fix the mass of dark vector $m_{\gamma'}=10^{-6}$~eV to establish the approximation of 
conversion probability, i.e., Eq.~\eqref{eq:prob4}. Note that our analysis is independent of $m_{\gamma'}$ as long as 
$m_{\gamma^{\prime}} \lesssim\left(T_s /R_{s}\right)^{1 / 2} \sim 3\times 10^{-5}~\rm{eV}$, with surface temperature $T_s\sim 50$~eV. 

The performance of the Bayesian statistical analysis is carried out using the {\tt UltraNest} package~\cite{Buchner2021}, 
which implements a nested sampling Monte Carlo technique~\cite{Skilling2004}.
It computes the (log-)likelihood as well as the marginal likelihood (``evidence'') $Z$ to perform model comparison. 
% and measure the prediction parsimony {\bf (What does this mean?)} of a model.
Meanwhile, the posterior probability distributions on model parameters are constructed to describe the parameter constraints of the data.
We assume a uniform prior distribution for the distance and a log-uniform prior distribution for $e^{\prime}$ and $\varepsilon$. We choose the number of live points
to be 20000, and the total number of the samples called in the fitting is 225802.

In each running, we simulate the NS cooling based on the modified {\tt NSCool} code that includes the dark vector emissivity. 
We then determine the surface luminosity as well as the surface temperature at the age of J1856.  
We obtain a minimum value (corresponding to the maximum value of the likelihood) of the reduced chi-square $\chi^2_{\rm min}/{\rm DOF}=1.02/3$ in the fitting to the data.
On the other hand, when the emissivity of dark vector is turned off in the fitting, we obtain a value of $\chi^2_{0}/{\rm DOF}=9.38/6$ with the NS standard cooling model.
We thus conclude that we have made the fitting much better by taking into account the emission of dark vectors.  Therefore, the data supports the existing of a dark vector with couplings summarized in Table~\ref{tab:fitResults}. 
%*************************************************************************
\begin{table} 
  \centering
  \caption{Summary of the fit results for the parameters.}
  \begin{tabular}{cccccccccccccc}
    \hline
    \hline
    % \toprule
  %   rd,dd,mS,mx,zta,ls,la,ks,cs,cx,w0,vtc,Tc,bH,Tn \\
      Parameter & Prior range & Best-fit & Mean/$1\sigma$ range & Median/$1\sigma$ range \\
    \hline
    % \midrule
    $d$ [kpc] & $[90,160]$ & 122.870 & $123.119/[110.550,135.688]$ & $123.124/[110.248,135.894]$ \\
    $e^{\prime}\times 10^{-15}$ & $[1.3,12.6]$ & $6.653$ & $5.559/[4.055,7.62]$ & $5.585/[4.083,7.430]$\\
    $\varepsilon \times 10^{-9}$ & $[0.1,20.0]$ & $0.641$ & $1.285/[0.478,3.459]$ & $1.282/[0.457,3.890]$\\
    \hline
    \hline
    % \bottomrule
  \end{tabular}
  \label{tab:fitResults}
\end{table} 
%*************************************************************************
% %*************************************************************************
% \begin{table} 
%   \centering
%   \caption{Summary of the fit results for the parameters.}
%   \begin{tabular}{cccccccccccccc}
%     \hline
%     \hline
%     % \toprule
%   %   rd,dd,mS,mx,zta,ls,la,ks,cs,cx,w0,vtc,Tc,bH,Tn \\
%       Parameter & Prior range & Best-fit & Mean$\pm$stdev & Median$\pm$error \\
%     \hline
%     % \midrule
%     $d$ [kpc] & $[90,160]$ & 122.870 & 123.119$\pm$12.569 & $123.124_{-12.876}^{+12.770}$ \\
%     ${\rm Log}_{10}e^{\prime}$ & $[-14.9,-13.9]$ & $-14.177$ & $-14.255\pm 0.137$ & $-14.253_{-0.136}^{+0.124}$\\
%     ${\rm Log}_{10}\varepsilon$ & $[-10.0,-7.7]$ & $-9.193$ & $-8.891\pm 0.430$ & $-8.892_{-0.448}^{+0.482}$\\
%     \hline
%     \hline
%     % \bottomrule
%   \end{tabular}
%   \label{tab:fitResults}
% \end{table} 
% %*************************************************************************

\begin{figure}
  %\centering
  \includegraphics[width=110mm,angle=0]{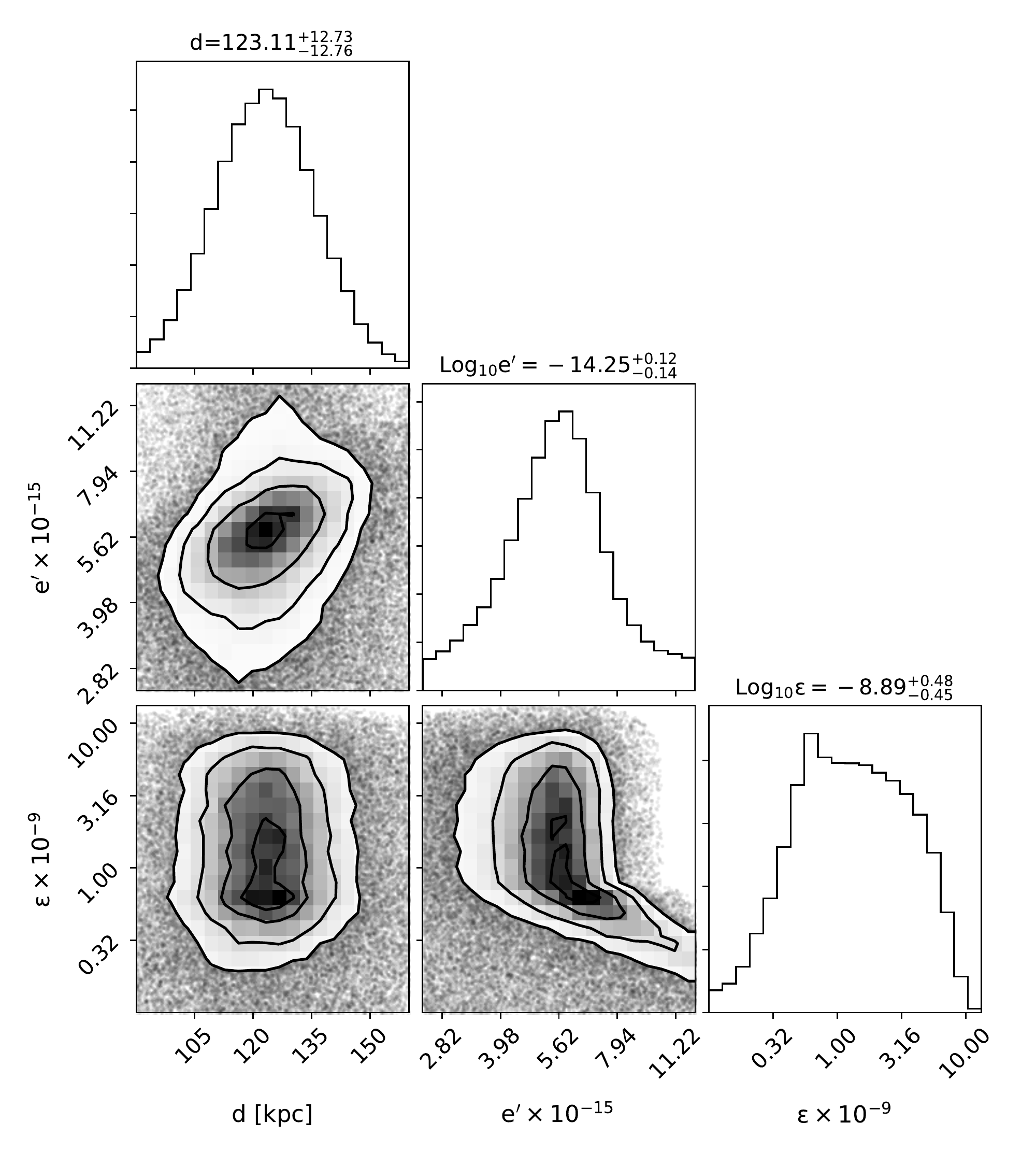}
  \caption{Corner plot of the posterior distributions of the parameters.}
  \label{fig:cornerplot}
\end{figure}

\begin{figure}
  %\centering
  \includegraphics[width=110mm,angle=0]{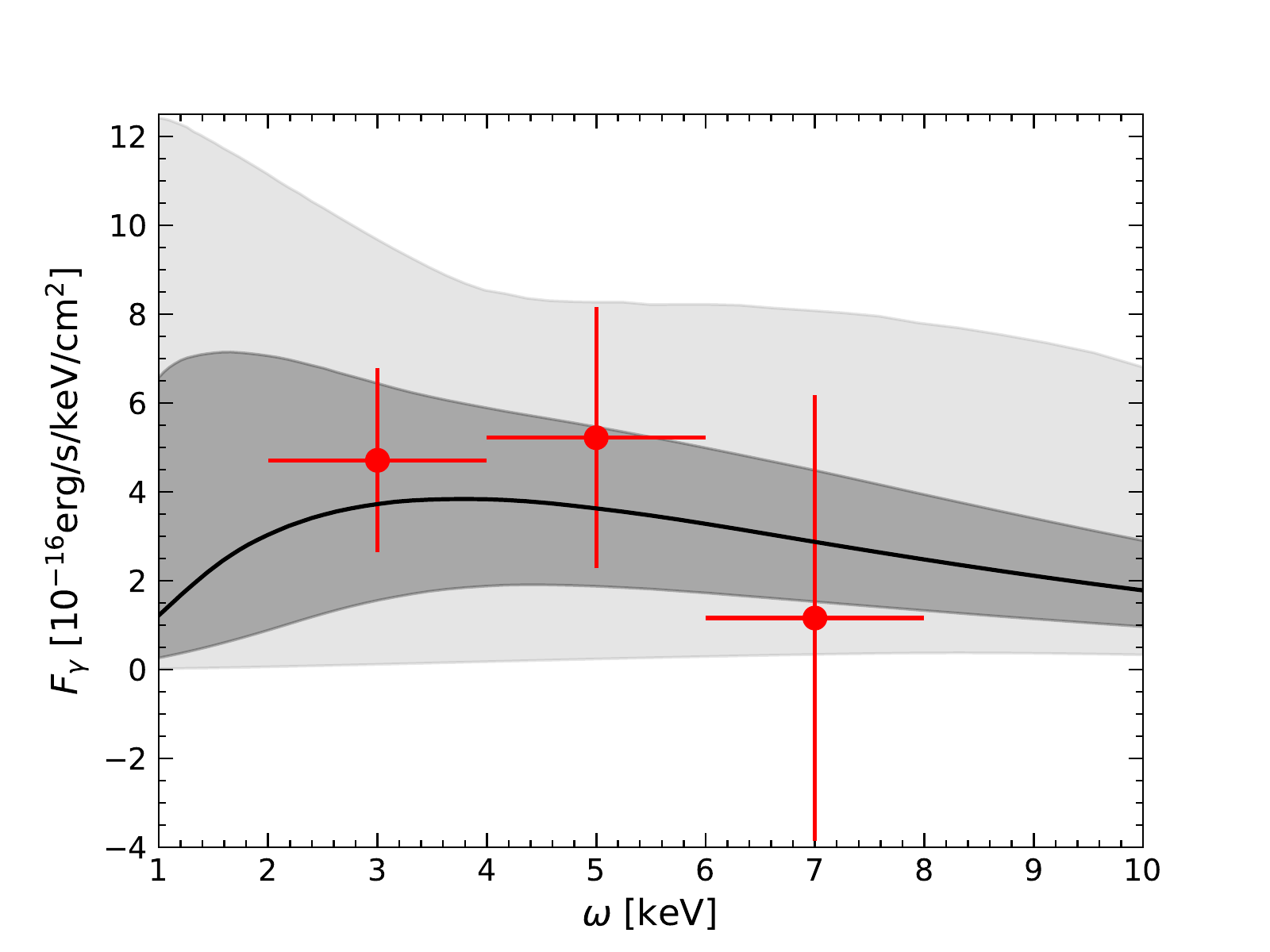}
  \caption{Bands of the model predictions for the spectrum as calculated from the chain.
  The black curve denotes the prediction of the model with the median values, the black band 
  represents the $1\sigma$~$(q=0.341)$ confidence interval, and the grey band represents the confidence interval with a quantile value $q=0.49$.
  The red points are the data of the J1856 x-ray spectrum.}
  \label{fig:bandplot}
\end{figure}

\begin{figure}
  %\centering
  \includegraphics[width=75mm,angle=0]{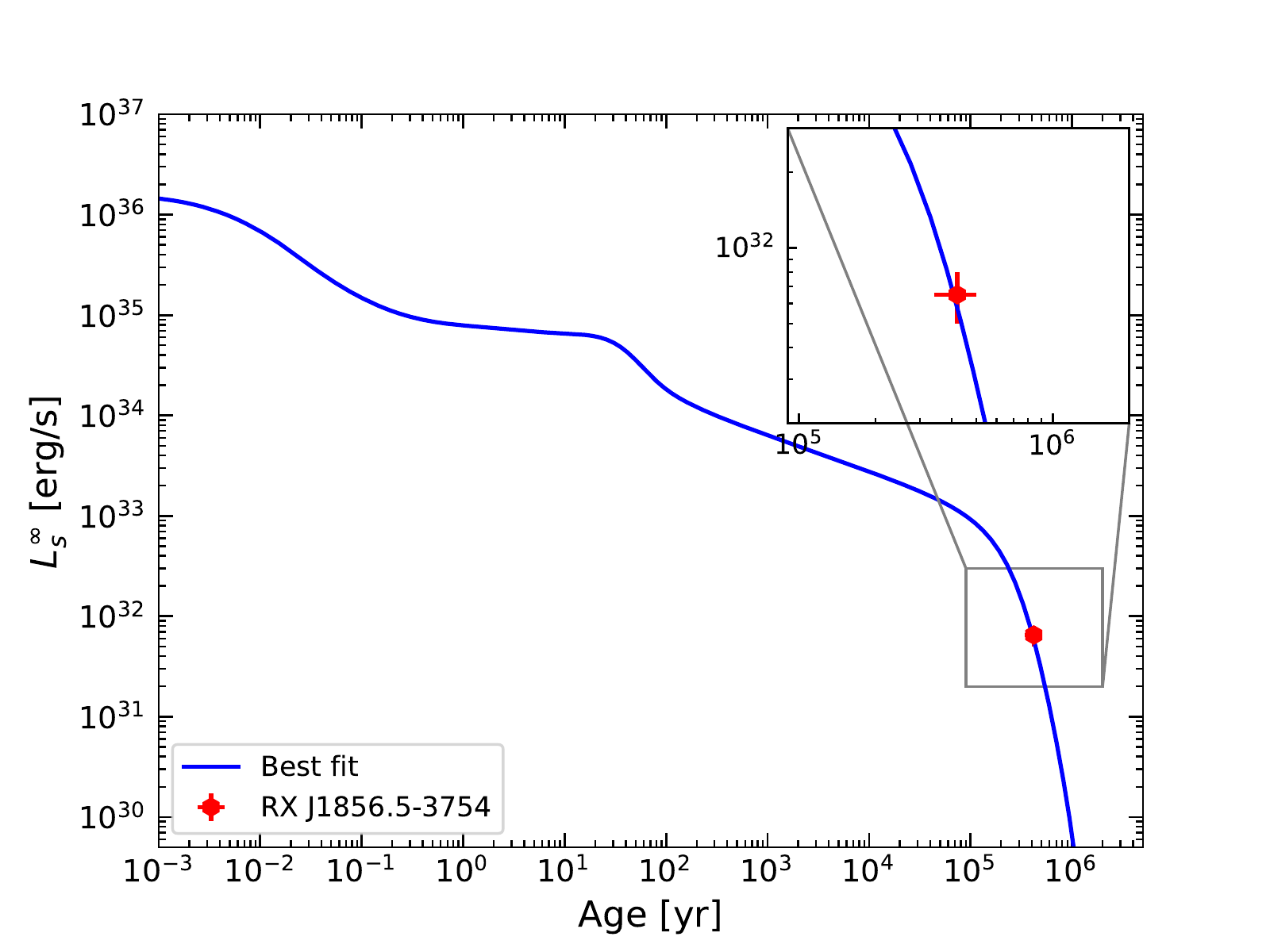}
  \includegraphics[width=75mm,angle=0]{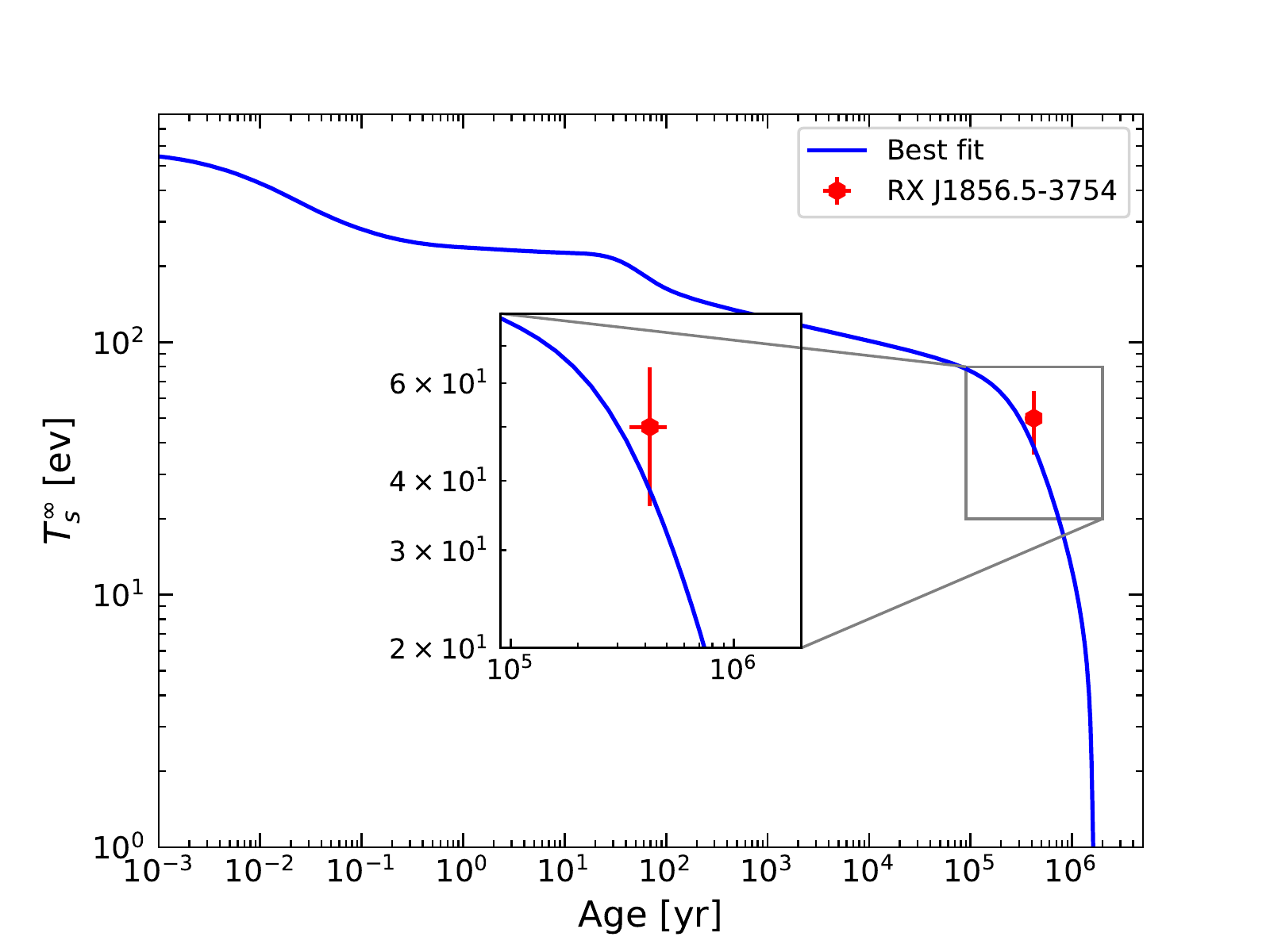}
  \caption{$L_{s}^{\infty}$ (left plot) and $T_s^{\infty}$ (right plot) as a function of age.  The blue curve represents the best-fit result and the red data point denotes the observation.}
  \label{fig:Best}
\end{figure}

The corner plot in Fig.~\ref{fig:cornerplot} shows the posterior distributions of the parameters. In Fig.~\ref{fig:bandplot}, we show the predictions of the model 
for the J1856 x-ray spectrum. The black curve denotes the prediction of the model with the median values, while the black band 
represents the $1\sigma$ confidence interval. Figure~\ref{fig:Best} depicts the evolution of the surface luminosity and temperature observed at infinity with the NS age.
In this plot, we take the parameters to their best-fit values. As indicated in these figures, the x-ray spectrum as well as the luminosity observations can be 
well interpreted in the $B-L$ dark vector model. We note that all of the PBF processes have been turned off in our simulations.  Including these processes would lead to much lower surface luminosities than the observations. 

%*****************************fig1***************************************
%*************************************************************************
%*****************************fig1***************************************
%*************************************************************************
%*****************************fig1***************************************
%*************************************************************************

\section{Model constraints}\label{sec:limits}

\begin{figure}
  %\centering
  \includegraphics[width=110mm,angle=0]{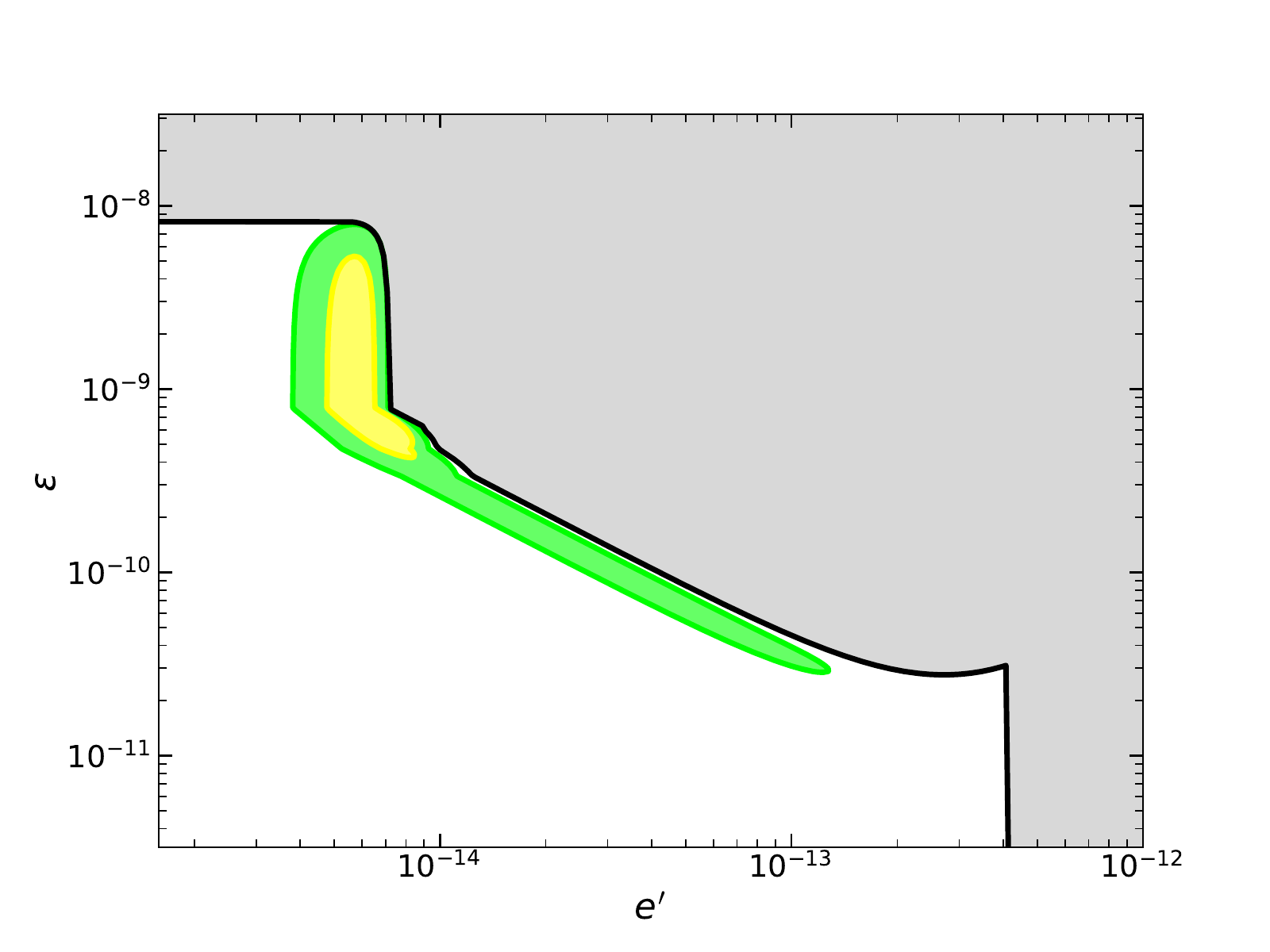}
  \caption{ The black curve denotes the constraint from J1856 observations at $2\sigma$ (95\%) confidence level, the grey region on 
  the $e^{\prime}-\varepsilon$ plane is excluded by the constraint. The green and yellow regions represent the parameter spaces that are favored
  by the observations at $1\sigma$ and $2\sigma$ confidence interval.}
  \label{fig:limit}
\end{figure}

The likelihood ratio test~\cite{Rolke2005} is used to determine the limit on and the significance of a possible dark vector contribution to the J1856 observations. 
To do this, we fix $e^{\prime}$ and determine the ``nuisance parameter'' $d$ by minimizing the chi-square. 
Upper limits at the 95\% confidence level on $\varepsilon$ are derived by increasing the chi-square from its minimum
value of the model until it changes by 2.71, $\chi_{\rm upp}^2(\varepsilon)=\chi_{\rm min}^2(\bar{\varepsilon})+2.71$, with $\bar{\varepsilon}$
denoting the parameter that minimizes the chi-square at fixed $e^{\prime}$. The black curve in Fig.~\ref{fig:limit} represents the constraints on
$e^{\prime}-\varepsilon$ and the grey region is excluded by the constraint. The green and yellow regions represent the parameter spaces that are favored
by the observations at $1\sigma$ and $2\sigma$ confidence intervals. The conclusions inferred from this figure are summarized as follows:
\begin{itemize}
  \item {There is an upper limit on the mixing angle $\varepsilon<7.97\times 10^{-9}$, which is independent of the gauge coupling $e^{\prime}$ 
  and dark vector mass $m_{\gamma^{\prime}}$ (as long as $m_{\gamma'}\lesssim 3\times 10^{-5}$~eV). 
  This is because the observed surface luminosity would be strongly suppressed by the $\gamma-\gamma'$ conversion when $\varepsilon\gtrsim 10^{-8}$, as 
  illustrated in Fig.~\ref{fig:epsLum}. This constraint can also be applied to the dark photon model.}
  \item {There is an upper limit on the gauge coupling $e^{\prime}<4.13\times 10^{-13}$, which is independent of the mixing angle $\varepsilon$ 
  and dark vector mass $m_{\gamma^{\prime}}$ (as long as $m_{\gamma'}\lesssim 1$~keV). This is because the surface luminosity would be strongly reduced due to the
  cooling of the NS by the dark vector emissivity with $e^{\prime}\gtrsim 5\times 10^{-13}$, as shown in Fig.~\ref{fig:lumComponent}.}
  \item {For the gauge coupling in the range of $6\times 10^{-15}\lesssim e^{\prime}\lesssim 10^{-13}$, the x-ray spectrum data dominates the contributions to
  chi-square. We have shown that the model with parameters around the mean values $e^{\prime}=5.56\times 10^{-15}$ and $\varepsilon=1.29\times 10^{-9}$ can 
  explain the x-ray spectrum excess in the energy band $2-8$~keV. Since the x-ray spectrum from the new physics is proportional to $(e^{\prime}\varepsilon)^2$,
  constraints on $\varepsilon$ become stronger with the increase of $e^{\prime}$.}
\end{itemize}

%*****************************fig1***************************************
%*************************************************************************

\begin{figure}
  %\centering
  \includegraphics[width=75mm,angle=0]{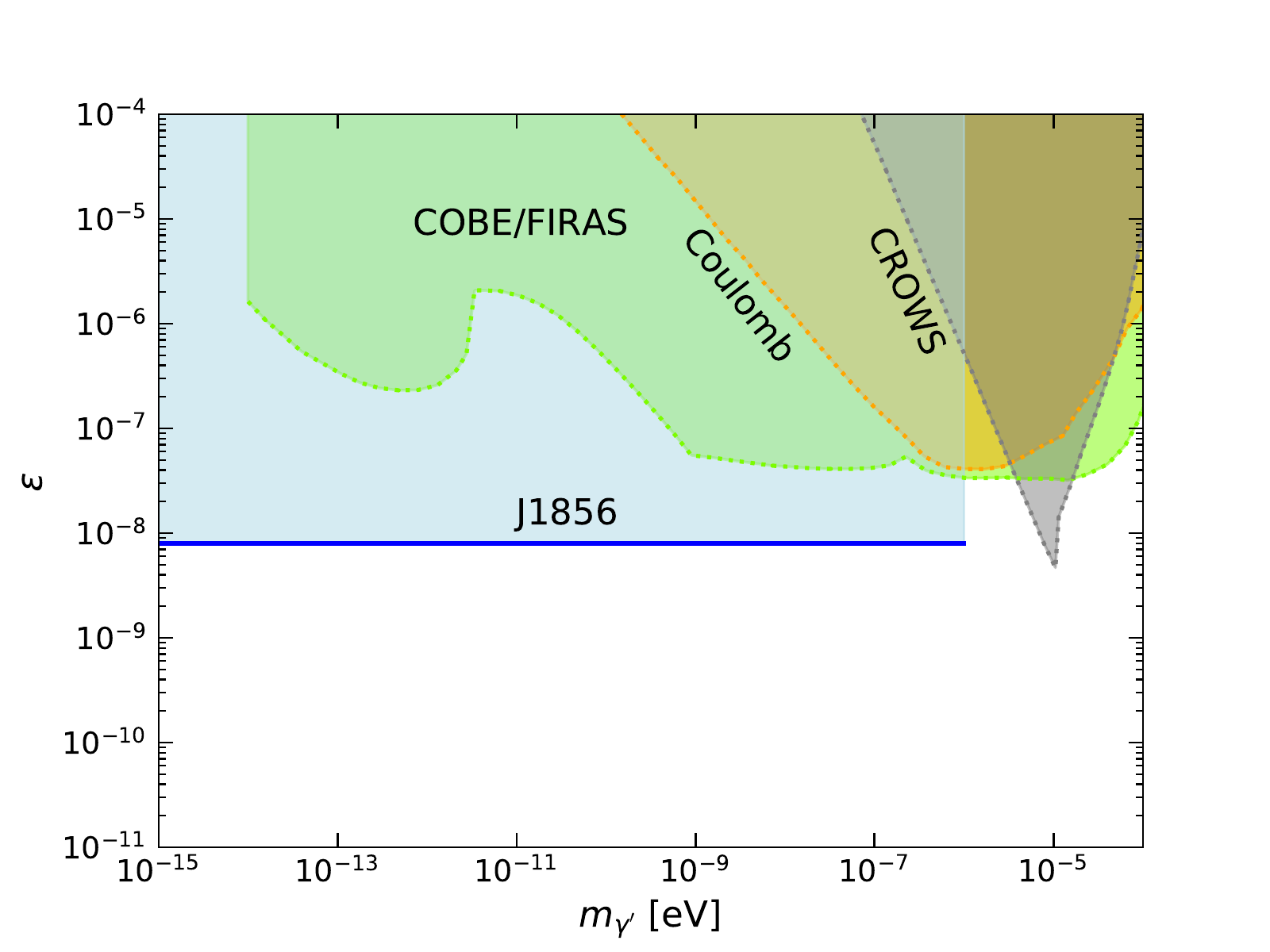}
  \includegraphics[width=75mm,angle=0]{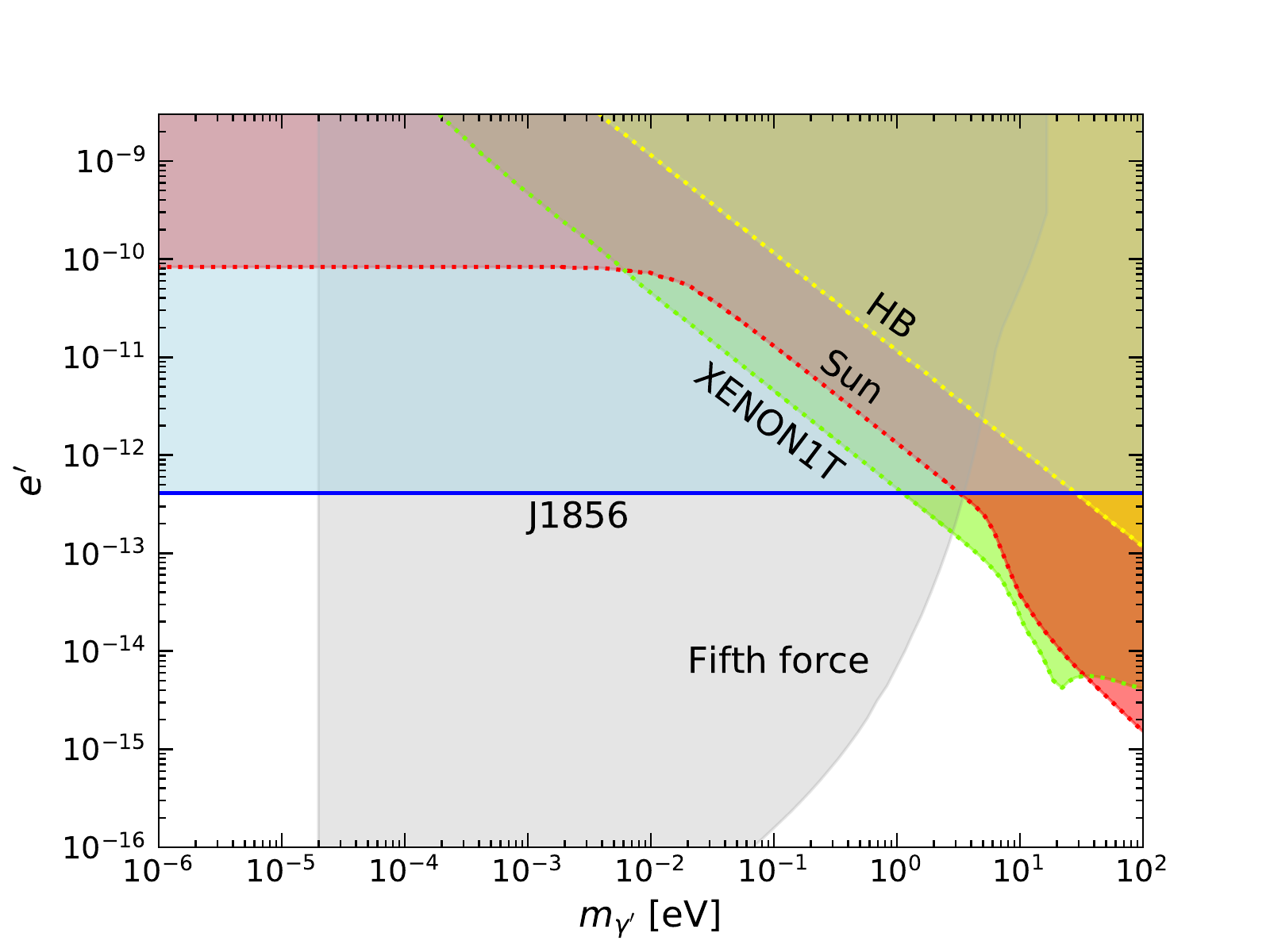}
  \caption{ Left: constraints on the mixing angle $\varepsilon$ as a function of $m_{\gamma'}$. 
  Our constraints from J1856 surface luminosity observations are depicted by the blue region.
  The light-green, orange, and grey regions denote the constraints from COBE/FIRAS~\cite{Fixsen1996APJ,Caputo2020PRL,Aramburo2020JCAP,McDermott2020PRD}, 
  Coulomb~\cite{Jaeckel2010PRD}, and CROWS~\cite{Betz2013PRD} experiments, respectively.
  Right: constraints on the gauge coupling $e'$ as a function of $m_{\gamma'}$. 
  The blue, red, orange, light-green, and light-grey regions represent the constraints from J1856 (our work), the Sun~\cite{Hardy2017JHEP}, 
  HB~\cite{Hardy2017JHEP}, XENON1T~\cite{An2020PRD}, and the fifth force experiments~\cite{Sushkov2011PRL,Murata2015CQG,Chen2016PRL}, respectively.
  The colored regions are excluded by the constraints.}
  \label{fig:limitCompare}
\end{figure}

In Fig.~\ref{fig:limitCompare}, we compare our constraints with those limits from cosmological, astrophysical, as well as terrestrial observations.
In the left plot of the figure, we summarize the constraints on the mixing angle $\varepsilon$ in the low mass range $\lesssim 10^{-4}$~eV.
The oscillation between the ordinary photon and the massive dark photon $\gamma\to \gamma^{\prime}$ induces deviations on the black body spectrum 
in the cosmic microwave background, which has been constrained by the COBE/FIRAS experiment~\cite{Fixsen1996APJ,Caputo2020PRL,Aramburo2020JCAP,McDermott2020PRD} (light-green region).
The constraints from detecting modifications of Coulomb force by the new gauge force in atomic and nuclear experiments are depicted by the orange region~\cite{Jaeckel2010PRD}.
Using the phenomenon of light shining through a wall for dark photons, the grey region has been bounded by the experiment CROWS~\cite{Betz2013PRD} at CERN. 
We observe that our constraint (light-blue region) on $\varepsilon$ from J1856 surface luminosity observation is much stronger than the above 
constraints by about an order of magnitude.

%*****************************fig1***************************************
%*************************************************************************
In the right plot of Fig.~\ref{fig:limitCompare}, we summarize the constraints on the gauge coupling $e^{\prime}$.
The exchange of light bosons, such as gauge bosons, scalar axions, and dilatons among others generates a Yukawa potential, 
which is tested by the fifth force experiments~\cite{Sushkov2011PRL,Murata2015CQG,Chen2016PRL}. Furthermore, light boson states emitted 
from the Sun may be probed in the DM direct detection experiments~\cite{An2020PRD,XENON1T2020,PandaX2021}. 
We compare our NS cooling constraint from J1856 (blue region) with the constraints given by the 
Sun (red)~\cite{Hardy2017JHEP}, HB (orange)~\cite{Hardy2017JHEP}, XENON1T (green)~\cite{An2020PRD}, and the current fifth force 
experiments (light-grey)~\cite{Sushkov2011PRL,Murata2015CQG,Chen2016PRL}. 
We observe that the fifth force experiments give the most stringent constraints for $2\times 10^{-5}~{\rm eV}\lesssim m_{\gamma^{\prime}}\lesssim 3$~eV.
Constraints from J1856 surface luminosity observation are much stronger than those given by the Sun luminosity observations and the DM direct detections,
for the dark vector with mass $\lesssim 1$~eV.
In summary, we conclude that the $2\sigma$ parameter space with $m_{\gamma^{\prime}}\lesssim 10^{-5}$~eV, for explaining the hard x-ray excess, remains available 
when various observation constraints on the dark vector model are taken into account.

%==================================================================================
\begin{table}[tbp]
  %\centering
  \renewcommand\arraystretch{1.5}
  \begin{tabular}{|c||c||c||c||c|c|}
  \hline
  Experiment & Energy band & Intensity limit & Predicted intensity & Reference \\
      %  & $10^{14}$~G & kyr & kpc & $10^{33}$~erg/s & \multicolumn{2}{c|}{}\\
  \hline 
  Swift	   & $14-195$ & $7\times 10^{-12}$   & $3.56\times 10^{-16}$  & \cite{Oh2018APJ}           \\
  \hline 
  INTEGRAL & $17-60$  & $1.3\times 10^{-11}$ & $1.74\times 10^{-16}$  & \cite{Krivonos2019MSAI}    \\
  \hline 
  NuSTAR   & $6-10$   & $3\times 10^{-15}$   & $1.03\times 10^{-15}$  & \cite{Buschmann2021PRL}    \\
  \hline
  NuSTAR   & $10-60$  & $2\times 10^{-14}$   & $8.59\times 10^{-16}$  & \cite{Buschmann2021PRL}    \\ 
  \hline 
  \end{tabular}
  \caption{\label{tab:prediction}The energy is in units of keV and the intensity is in units of $\rm erg/cm^{2}/s$.
  The third column shows the current limit (Swift and INTEGRAL) and future sensitivity (NuSTAR) on the x-ray intensity, 
  whereas the fourth column lists the predicted intensities, assuming the median values of the parameters for the dark vector model.}
\end{table}
%==================================================================================

In Table~\ref{tab:prediction} we summarize the current limits on hard x-ray intensity for NSs near the galactic plane (J1856, J0806, J0720, and J2143) from the 105-month Swift Burst Alert Telescope all-sky hard x-ray $|b| \leq 17.5^{\circ}$ survey~\cite{Oh2018APJ} and the 14-year INTEGRAL galactic plane 
survey~\cite{Krivonos2019MSAI}. We also adopt the projected sensitivity (provided in the supplementary materials of Ref.~\cite{Buschmann2021PRL}) 
at 95\% confidence level for a 400~ks NuSTAR observation of J1856 in two energy bands. The third column provides the predicted intensities, assuming the median values of the parameters for the dark vector model. We observe that the predicted intensities are far below the current limits from Swift and INTEGRAL, and future observations on M7 by the NuSTAR experiment may be useful to test the dark vector model.

%%%%%%%%%%%%%%%%%%%%%%%%%%%%%%%%%%%%%%%%%%%%%%%%%%
\section{Summary and conclusions}
\label{sec:conclusion}
%%%%%%%%%%%%%%%%%%%%%%%%%%%%%%%%%%%%%%%%%%%%%%%%%%

The NS is recognized as one of the most excellent astrophysical laboratories for searching for new light particles that couple weakly to SM particles.
In this work, we have presented the emission of light dark vectors from the nucleon bremsstrahlung processes in the core of the NS.
The dark vector is assumed to be a $B-L$ gauge boson and to have a mass much below about keV, the core temperature of the NS.
The dominant production mode of dark vector in the NS core is the neutron bremsstrahlung since the production of dark vector by the charged particles in the NS core 
is suppressed by a factor of $m_{\gamma'}^2$ from the in-medium effect.
Since the plasma is thought to be dilute and nonrelativistic in the Sun and supernova, the nondegenerate limit was employed to 
determine the dark vector emissivity and obtain constraints in previous literature.  
In the current work, we present the calculation of the emission rate of dark vectors from the nucleon bremsstrahlung processes in the degenerate limit, 
which is the case for the strongly-compressed circumstances in the NS core.
In addition, we also calculate the photon luminosity observed at infinity by taking into account the photon-dark vector conversion during their propagation.

In this work, we attempt to interpret the J1856 hard x-ray spectrum excess in terms of the dark vector model while taking into account the J1856 surface luminosity and temperature observations. 
The thermal photon spectrum makes negligible contribution to the hard x-ray observations since the energy of the thermal spectrum peak is determined by the 
surface temperature $T_s\sim 50$~eV. On the other hand, the peak of the spectrum from $\gamma^{\prime}\to \gamma$ conversion is obtained at a higher energy that is determined
by the core temperature $T_c\sim 2$~keV.  Therefore, we expect that the dark vector emission can lead to the hard x-ray excess.
The evolution of the NS with time would be altered if its energy loss was dominated by the production of dark vectors rather than the standard neutrino emission. 
We perform numerical simulations of the NS cooling based on the modified {\tt NSCool} code that includes additional energy loss via the dark vector emission, assuming the APR EOS for the NS core with a canonical mass of $1.4~M_{\odot}$.
In this way, we determine the surface luminosity, as well as the surface temperature, for the NS with a given age. 
Then we calculate the hard x-ray spectrum from $\gamma^{\prime}\to \gamma$ conversion and consider the inverse conversion for the surface luminosity observed at infinity.
We carry out the Bayesian statistical analysis of the J1856 data using the {\tt UltraNest} package and employ the likelihood ratio test to construct 95\% confidence level upper limits on the parameters of the dark vector model.
Our findings are summarized as follows:
\begin{itemize}
  \item {The fit to the J1856 hard x-ray excess data favors the dark vector model with $e^{\prime}=5.56\times 10^{-15}$ and $\varepsilon=1.29\times 10^{-9}$.
  Furthermore, the $2\sigma$ parameter space with $m_{\gamma^{\prime}}\lesssim 10^{-5}$~eV for the interpretation of the hard x-ray excess does not conflict 
  with any of the currently observed constraints.}
  \item {Due to the $\gamma\to\gamma^{\prime}$ conversion, there exists an upper limit on the mixing 
  angle, $\varepsilon<7.97\times 10^{-9}$, for $m_{\gamma'}\lesssim (T_s/R_s)^{1/2}\sim 3\times 10^{-5}$~eV from the J1856 surface luminosity observation. This constraint is independent of the production of the dark vectors, and therefore, is independent of the gauge coupling $e^{\prime}$,
  and thus can be applied to the dark photon model.}
  \item {The emission of the dark vectors accelerates the NS cooling and reduces the surface luminosity of J1856, which leads to an upper limit on the gauge 
  coupling, $e^{\prime}<4.13\times 10^{-13}$, at 95\% confidence level for $m_{\gamma'}\lesssim 1$~keV.}
\end{itemize}

Our best-fit dark vector model predicts much lower hard x-ray intensities than the current limits from the Swift and INTEGRAL hard x-ray surveys.
Future hard x-ray observations of J1856 by NuSTAR in particular may constrain or provide additional evidence for the best-fit dark vector from this work. 
NSs are promising targets for testing the weakly-interacting light dark vector particle. The constraints on the dark vector model can be further improved 
with more hard x-ray observations of the NSs from future telescopes.

\acknowledgments{ BQL thanks Lev Leinson a lot for helpful suggestions on the NSCool code.
This work was supported in part by the Ministry of Science and Technology (MOST) of Taiwan under Grant Nos.~MOST-108-2112-M-002-005-MY3 and 109-2811-M-002-550. 
}

\appendix

%%%%%%%%%%%%%%%%%%%%%%%%%%%%%%%%%%%%%%%%%%%%%%%%%%
\section{In-medium effects for (dark) photons}
\label{adp:imf}
%%%%%%%%%%%%%%%%%%%%%%%%%%%%%%%%%%%%%%%%%%%%%%%%%%

This Appendix briefly reviews the results of photon self-energies in plasmas given in Refs.~\cite{Braaten1993PRD,Blaizot2001PRD}.
To leading order in the EM coupling constant $\alpha$, the EM polarization tensor $\Pi^{\mu \nu}$ that determines
the effects of a plasma on the propagation of photons is given by~\cite{Braaten1993PRD}
\begin{equation}
  \Pi^{\mu \nu}(K)=16 \pi \alpha \int \frac{d^{3} p}{(2 \pi)^{3}}\frac{f_e+f_{\bar{e}}}{2E}
  \frac{P \cdot K\left(P^{\mu} K^{\nu}+K^{\mu} P^{\nu}\right)-K^{2} P^{\mu} P^{\nu}-(P \cdot K)^{2} g^{\mu \nu}}{(P \cdot K)^{2}
  -\left(K^{2}\right)^{2}/4},
\end{equation}
where $f_e$ and $f_{\bar{e}}$ are the Fermi distribution function for electron and positron, $E=\sqrt{p^2+m_e^2}$,
$K^{\mu}=(\omega, \mathbf{k})$, $P^{\mu}=(E, \mathbf{p})$, $K^{2}=\omega^{2}-k^{2}$, and $P \cdot K=E \omega-\mathbf{p} \cdot \mathbf{k}$.
The integration over the angular parts can be performed by ignoring the $K^4$ term.  Taking into account the degenerate $\left(T \ll \mu-m_{e}\right)$
and relativistic $\left(T \gg m_{e}\right.$ or $\left.\mu \gg m_{e}\right)$ limits, one obtains the transverse and longitudinal polarization functions,
\begin{eqnarray}
  \Pi_{\mathrm{T}}&=&\omega_p^{2}\left[1+\frac{1}{2} G\left(v_{*}^{2} k^{2} / \omega^{2}\right)\right] \equiv \pi_{\mathrm{T}},\\
  \Pi_{\mathrm{L}}&=&\omega_p^{2} \frac{k^{2}}{\omega^{2}} \frac{1-G\left(v_{*}^{2} k^{2}/\omega^{2}\right)}{1-v_{*}^{2}k^{2}/\omega^{2}} 
  \equiv \frac{k^{2}}{\omega^{2}-k^{2}} \pi_{\mathrm{L}},
\end{eqnarray}
where $v_*$ denotes the typical electron velocity in the plasma, $\omega_p$ is the plasma frequency, which is dominated by the electrons
\begin{equation}
  \omega_{p}=\left(\frac{4 \pi \alpha n_{e}}{E_{F, e}}\right)^{1/2}~{\rm with~}E_{F,e}^{2}=m_{e}^{2}+\left(3 \pi^{2} n_{e}\right)^{2/3}.
\end{equation}
Under the high density circumstance in the NS core, the electron Fermi momentum $p_{F,e}=\left(3\pi^{2}n_{e}\right)^{1/3}\sim \mathcal{O}(100)$~MeV$\gg m_e,~T$. 
Finally, the function,
\begin{equation}
  G(x)=\frac{3}{x}\left(1-\frac{2 x}{3}-\frac{1-x}{2 \sqrt{x}} \ln \frac{1+\sqrt{x}}{1-\sqrt{x}}\right).
\end{equation}

In the Coulomb gauge, the transverse and longitudinal components of the effective propagator for the EM field are given by
\begin{eqnarray}
  D_{\rm T}^{i,j}(\omega,k)&=&\frac{1}{\omega^{2}-k^{2}-\Pi_{\mathrm{T}}}\left(\delta^{i j}-\frac{k^{i} k^{j}}{k^{2}}\right),\\
  D_{\rm L}^{0,0}(\omega,k)&=&\frac{1}{k^{2}-\Pi_{\mathrm{L}}}.
\end{eqnarray}
With the explicit expression of the photon propagator, one can find that the emission of dark vectors is given by
the vacuum matrix element for the emission of massive photons and multiplied by the effective coupling given by 
Eq.~\eqref{eq:eeff}~\cite{Hong2020}, {\it i.e.,}
\begin{equation}
  \label{eq:matrixinmed}
  \mathcal{M}_{\rm T,L}=e_{\mathrm{eff}}^{f}J_{f, \mu} \epsilon_{\mathrm{T}, \mathrm{L}}^{\mu}.
\end{equation}
This equation shows that the dark vector produced by the EM charged current is suppressed by the dark vector mass. While
for neutral currents, the effective coupling $e_{eff}^n=e'$, thus the dark vector produced from this process in medium is the same as that 
in the vacuum and the production is not suppressed~\cite{Hong2020}.

%%%%%%%%%%%%%%%%%%%%%%%%%%%%%%%%%%%%%%%%%%%%%%%%%%
\section{Conversion probability}
\label{apd:prob}
%%%%%%%%%%%%%%%%%%%%%%%%%%%%%%%%%%%%%%%%%%%%%%%%%%

The dark vectors emitted from the neutron bremsstrahlung processes may be converted into x-ray photons as they propagate outwards through 
the magnetosphere around the magnetized NS.  The stellar magnetic field is assumed to be dipolar 
\begin{equation}
  B(r) = B_{0}\left(\frac{r_{0}}{r}\right)^{3},
\end{equation}
where $B_0$ is the magnetic field at the surface of the NS, and $r_0$ is the NS radius.
% Under the approximation where we assume all axions travel along radial trajectories originating from the MWD center, 
% we may derive a simple analytic expression for the conversion probability.
% For both the perpendicular modes and the parallel modes, 
The evolution of the photon and the dark vector in the presence of an external magnetic field can be described in terms of the following system of 
first-order differential equations~\cite{Raffelt1988PRD,Fortin2019JCAP}
\begin{equation}
  \label{eq:nonQED}
  \left[i \partial_{r}+\omega+\left(\begin{array}{cc}
  \Delta & \varepsilon \Delta \\
  \varepsilon \Delta & \Delta_{\gamma'}
  \end{array}\right)\right]\left(\begin{array}{c}
  A \\
  A^{\prime}
  \end{array}\right)=0,
\end{equation}
where the term $\Delta_{\gamma'}=-\frac{m_{\gamma'}^{2}}{2 \omega}$ ($\omega$ being the energy of the fields) is due to the finite dark vector mass. 
The condition $m_{\gamma'}\ll \omega$ is required to satisfy the weak-dispersion limit~\cite{Lai2006PRD}.
These equations can describe the propagations of both the perpendicular modes and the parallel modes. 
Strong-field QED effects in vacuum give rise to the term~\cite{Heyl1997JPA}
\begin{equation}
  \Delta=\frac{1}{2}q\omega \sin^2\theta,
\end{equation}
where $\theta$ is the angle between the direction of propagation and the magnetic field, and $q$ is a 
dimensionless function of $b=B(r)/B_c$ (where the critical QED field strength $B_{c} \equiv m_{e}^{2} c^{3} /(e \hbar)=4.414 \times 10^{13}~\mathrm{G}$)
given by~\cite{Raffelt1988PRD,Potekhin2004APJ}
\begin{eqnarray}
  \label{eq:qper}
  q_{\perp }&=&\frac{4 \alpha}{45 \pi} b^{2} \hat{q}_{\perp }, ~ \text{ with } ~ \hat{q}_{\perp }=\frac{1}{1+0.72 b^{1.25}+0.27 b^{2}},\\
  \label{eq:qpar}
  q_{\parallel }&=&\frac{7 \alpha}{45 \pi} b^{2} \hat{q}_{\parallel }, ~ \text{ with } ~ \hat{q}_{\parallel }=\frac{1+1.25 b}{1+1.33 b+0.56 b^{2}},
\end{eqnarray}
where $\perp$ and $\parallel$ denote the perpendicular and parallel modes of the dark vectors, respectively. 
These formulae have the correct $b\gg 1$ and $b\ll 1$ limits. Since the observer is far away from the source, we take the latter limit, 
in which we have $\hat{q}_{\perp/{\parallel }}\simeq 1$. Note that these results are only valid for photon energies below the electron mass $m_e\sim 500$~keV, 
which is applicable for dark vector photon with energies in the hard x-ray frequency regime.

We follow Ref.~\cite{Raffelt1988PRD} to resolve differential equations~\eqref{eq:nonQED} in the weak-mixing limit.
Equation~\eqref{eq:nonQED} can be rewritten as a ``Schr{\"o}dinger equation'':
\begin{equation}
  i \partial_{r} \mathbf{A}=\left(\mathcal{H}_{0}+\mathcal{H}_{1}\right) \mathbf{A}
  ~,
\end{equation}
where $\mathbf{A}^{\top }=\left(A,~A'\right)$ and the ``Hamiltonian'' matrices are
\begin{equation}
  \mathcal{H}_{0}=\begin{pmatrix}
    \omega+\Delta & 0\\ 
    0 & \omega+\Delta_{\gamma^{\prime}}
  \end{pmatrix}~~{\rm and}~~
    \mathcal{H}_{1}=\begin{pmatrix}
      0 & \varepsilon \Delta \\ 
      \varepsilon \Delta & 0
  \end{pmatrix}.
\end{equation}
The Schr{\"o}dinger equation in the interaction representation is given by
\begin{equation}
  i \partial_{r} \mathbf{A}_{\text {int }}=\mathcal{H}_{\text {int }} \mathbf{A}_{\text {int}}
  ~,
\end{equation}
where 
\begin{equation}
  \label{eq:tf}
  \mathbf{A}_{\text {int }}=\mathcal{U}^{\dagger} \mathbf{A}~~{\rm and}~~
  \mathcal{H}_{\text {int }}=\mathcal{U}^{\dagger} \mathcal{H}_{1} \mathcal{U}
  ~.
\end{equation}
The ``transformation operator'' is defined as
\begin{equation}
  \mathcal{U}(r)=\exp \left(-i \int_{r_0}^{r} \mathcal{H}_{0}\left(r^{\prime}\right) d r^{\prime}\right),
\end{equation}
where $r_0$ is the initial position.
% The solution $\mathbf{A}_{\mathrm{int}}=\sum_{n}a_n\mathbf{A}_{\mathrm{int}}^{n}$, with order $n=0,1,2...$. 
The solution at order $n+1$ can be obtained order by order:
\begin{equation}
  \mathbf{A}_{\mathrm{int}}^{n+1}(r)=-i \int_{r_0}^{r} d r^{\prime} \mathcal{H}_{\mathrm{int}}\left(r^{\prime}\right) 
  \mathbf{A}_{\mathrm{int}}^{n}\left(r^{\prime}\right),
\end{equation}
with the zero-order solution $\mathbf{A}_{\text {int }}^{0}(r)=\mathbf{A}(r_0)$. 
For the first-order solution, we have 
\begin{equation}
  \mathbf{A}_{\mathrm{int}}^{1}(r)=-i \mathbf{M}(r)\mathbf{A}(r_0),
\end{equation}
where the matrix
\begin{equation}
  \mathbf{M}(r)=\begin{pmatrix}
    0 & \int_{r_0}^{r}\varepsilon \Delta(r') e^{i\left[ I_1(r')-I_2(r') \right]}dr^{\prime} \\ 
    \int_{r_0}^{r}\varepsilon \Delta(r') e^{i\left[ I_2(r')-I_1(r') \right]}dr^{\prime}  & 0
  \end{pmatrix},
\end{equation}
with
\begin{equation}
  I_1(r')=\int_{r_0}^{r'}\Delta(r'') dr''~~{\rm and}~~I_2(r')=\int_{r_0}^{r'}\Delta_{\gamma^{\prime}} dr''.
\end{equation}
The dark vector-photon conversion probability at the first-order (in the interaction representation) is 
\begin{equation}
  \label{eq:intCP}
  P_{\gamma' \rightarrow \gamma}=|\mathbf{M}_{12}|^2
  =\varepsilon^{2}\left|\int_{r_0}^{r}dr^{\prime} \Delta\left(r^{\prime}\right) 
  \exp \left\{i \int_{r_0}^{r^{\prime}}dr^{\prime \prime}\left[\Delta\left(r^{\prime \prime}\right)-\Delta_{\gamma'}
  \right]\right\}\right|^{2}.
\end{equation}
Obviously, the conversion probability in the Schr$\ddot{\rm o}$dinger representation equals to Eq.~\eqref{eq:intCP} with transformations~\eqref{eq:tf}.

%%%%%%%%%%%%%%%%%%%%%%%%%%%%%%%%%%%%%%%%%%%%%%%%%%
\section{Analytical results for conversion probability}
\label{apd:analyticalProb}
%%%%%%%%%%%%%%%%%%%%%%%%%%%%%%%%%%%%%%%%%%%%%%%%%%

\begin{figure}
  %\centering
  \includegraphics[width=75mm,angle=0]{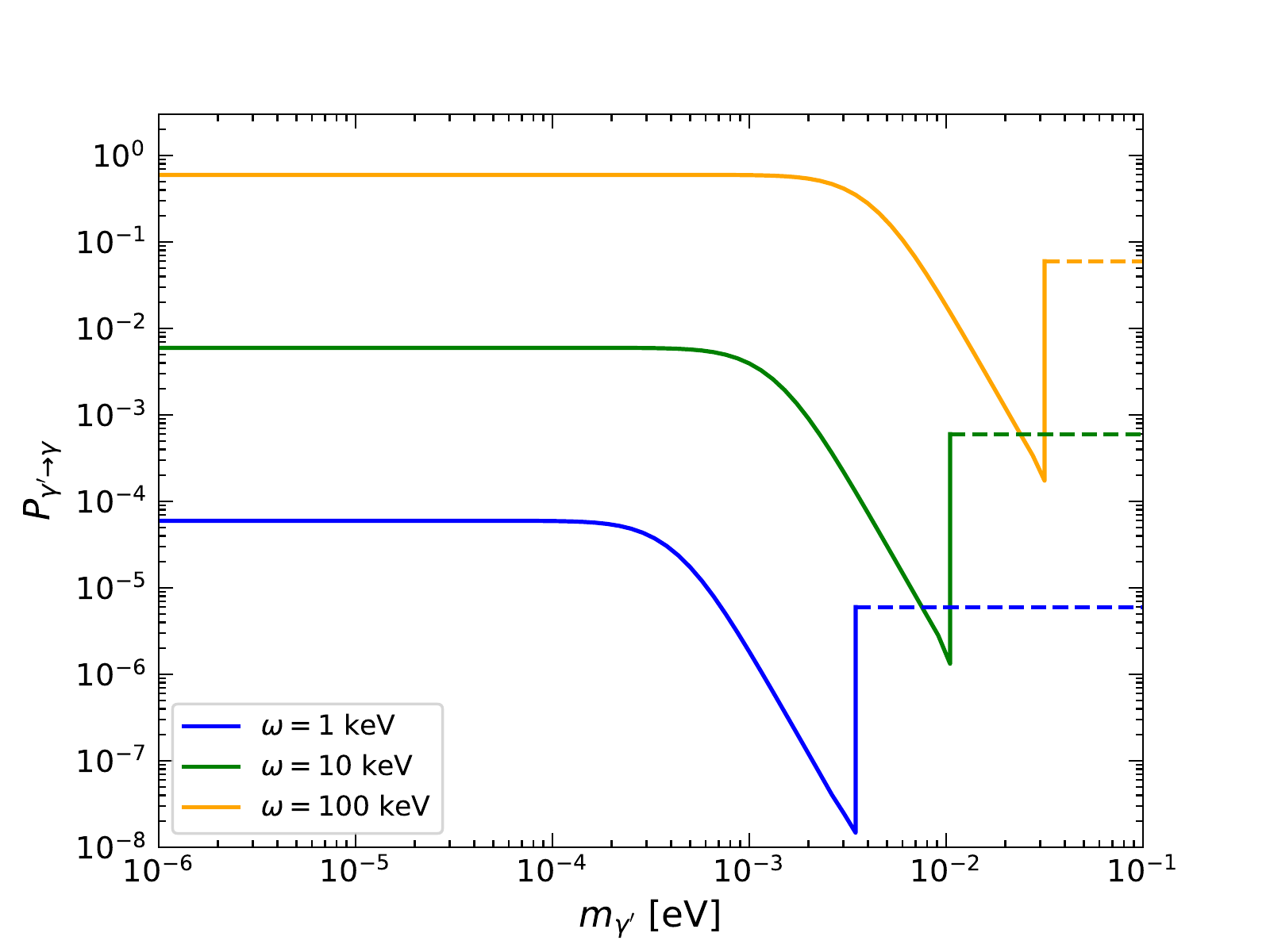}
  \includegraphics[width=75mm,angle=0]{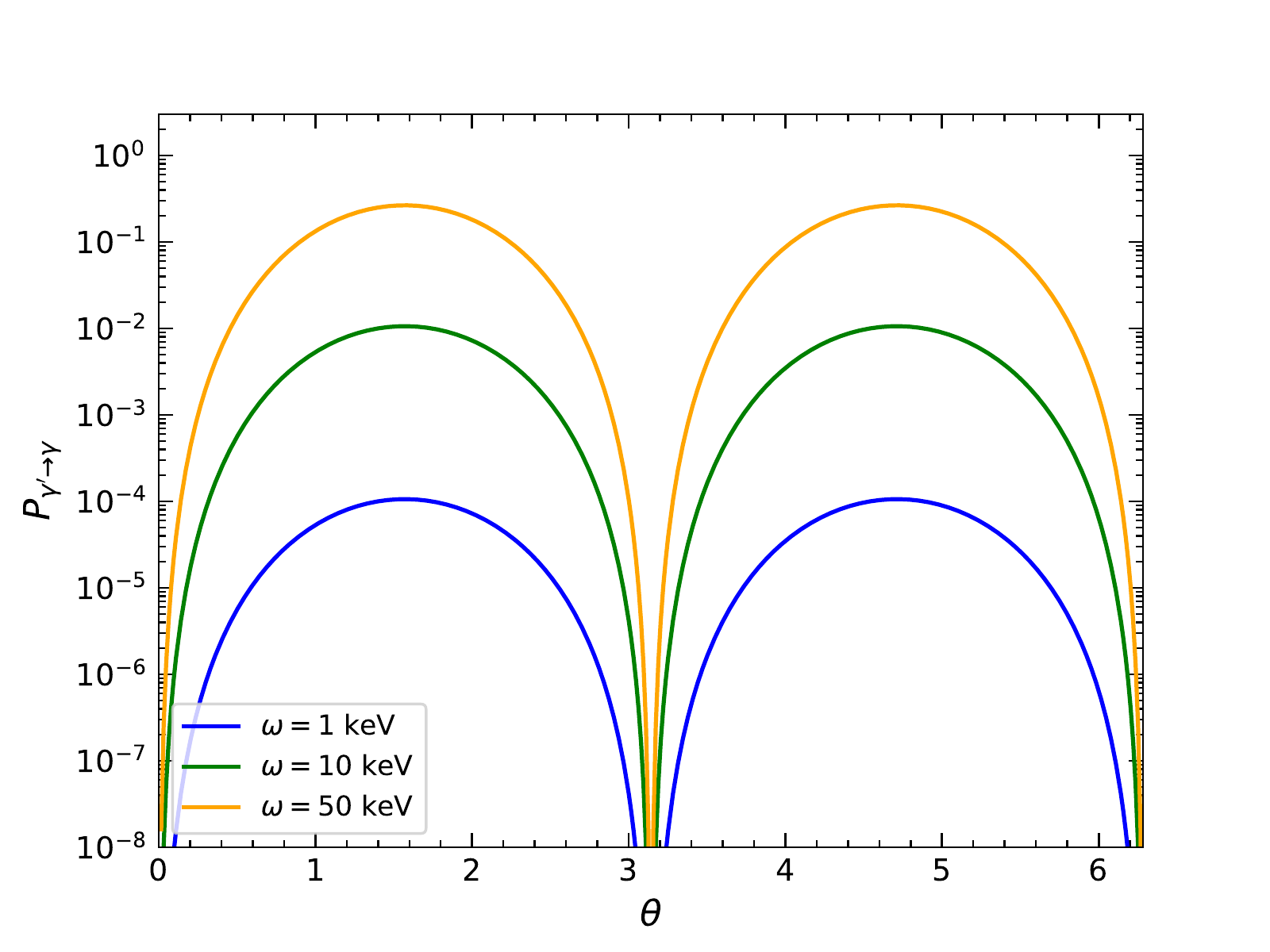}
  \caption{Left: the conversion probability $P_{\gamma'\to\gamma}$ as a function of $m_{\gamma'}$, with $\theta=\pi/3$.
  The dashed curves denote the constants at which the probabilities oscillate around.
  Right: the conversion probability $P_{\gamma'\to\gamma}$ as a function of $\theta$, with $m_{\gamma'}=10^{-5}$~eV.
  In both plots, we assume $B_0=10^{14}$~G and $\varepsilon=10^{-12}$.}
  \label{fig:prob}
\end{figure}

In this section, we show some analytical results for the dark vector-photon conversion probability to directly see how the probability 
is enhanced by a strong magnetic field. Since both modes of dark vector obey the same equations of motion~\eqref{eq:nonQED}, 
we will focus on the parallel mode below. 
The dark vector-photon conversion probability in the weak-mixing limit is given by 
\begin{eqnarray}
    \label{eq:prob2}
    P_{\gamma' \rightarrow \gamma}
    &=&\varepsilon^{2}\left|\int_{r_0}^{r}dr^{\prime} \Delta_{\parallel }\left(r^{\prime}\right) 
    \exp \left\{i \int_{r_0}^{r^{\prime}}dr^{\prime \prime}\left[\Delta_{\parallel }\left(r^{\prime \prime}\right)-\Delta_{\gamma'}
    \right]\right\}\right|^{2}\nonumber\\
    &\simeq &\varepsilon^{2}\left|\int_{r_0}^{r}dr^{\prime} \Delta_{\parallel }\left(r^{\prime}\right) 
    \exp \left\{i \left[ \left( \Delta_{\gamma'}+y/10 \right)r_0-\Delta_{\gamma'}r'\right]\right\}\right|^{2},
\end{eqnarray}
where $y=\frac{7\alpha}{45\pi}\left(\frac{B_0}{B_c}\right)^2\omega \sin^2\theta$. 
% The approximation $\hat{q}_{\parallel}\simeq 1$ has been used here.
We plot $P_{\gamma' \rightarrow \gamma}$ as a function of 
$m_{\gamma'}$ and $\theta$ in Fig.~\ref{fig:prob}. In both plots, we take $B_0=10^{14}$~G and $\varepsilon=10^{-12}$.
As shown in the left plot of Fig.~\ref{fig:prob}, the probability $P_{\gamma' \rightarrow \gamma}$ is a constant 
in the limit of low dark vector mass, $m_{\gamma'}\lesssim (\omega/R_s)^{1/2}\sim 10^{-4}$~eV at frequencies $\omega\sim 1$~keV and NS 
radius $R_s\simeq 11$~km. Due to the term $\Delta_{\gamma^{\prime}} r^{\prime}$ in the exponent of Eq.~\eqref{eq:prob2}, as $m_{\gamma'}$ further increases, 
the probability is suppressed and oscillates around a constant, which is represented by the dashed line in the left plot.
In the right plot we show the probability as a function of $\theta$, the angle between the direction of propagation and the magnetic field, with $m_{\gamma'}=10^{-5}$~eV.
% We fix $m_{\gamma'}=10^{-5}$~eV for the right plot of Fig.~\ref{fig:prob}. 
% The probability increases with $\omega$ and is proportional to $\sin^4\theta$. 
% In our calculation of x-ray flux, we have averaged over $\theta$, the angle between the direction of propagation and the magnetic field.
%*****************************fig1***************************************
%*************************************************************************

For small values of $m_{\gamma^{\prime}}$, the oscillation term $\Delta_{\gamma^{\prime}} r^{\prime}$ can be neglected and the conversion probability is approximated as
\begin{equation}
  \label{eq:prob3}
  P_{\gamma' \rightarrow \gamma}\simeq \left[\frac{7\alpha}{450\pi}\left(\frac{B_0}{B_c}\right)^2\varepsilon\omega r_0\sin^2\theta \right]^2.
\end{equation}
Averaged over $\theta$, the probability can be parametrized as
\begin{equation}
  \label{eq:prob4}
  \bar{P}_{\gamma' \rightarrow \gamma}=1.49\times 10^{-5}\left(\frac{\varepsilon}{10^{-12}}\right)^2\left( \frac{B_0}{10^{14}~{\rm G}}\right)^4
  \left(\frac{r_0}{11~{\rm km}} \right)^2\left( \frac{\omega}{1~{\rm keV}}\right)^2.
\end{equation}
Note that we have used the $\theta$-averaged probability in our calculations for the luminosity and spectrum.
In the presence of an inhomogeneous external field, the probability is proportional to $\omega^2$ for the dark vector-to-photon 
conversion~\cite{Fortin2019JCAP}, but is inversely proportional to the frequency in the zero background field 
limit~\cite{Masso2006PRL,Ahlers2007PRD,Ahlers2008PRD}.
Furthermore, when the external field is removed, the conversion probability approaches zero in the limit of low dark vector mass since in 
this case the probability is proportional to $m_{\gamma'}^2$~\cite{Masso2006PRL,Ahlers2007PRD,Ahlers2008PRD}.  However, for the case of an
inhomogeneous external field, the dark vector-to-photon conversion probability dose not depend on $m_{\gamma'}$ in the low dark vector mass 
limit~\cite{Fortin2019JCAP}, which also appears in the axion-photon conversion~\cite{Buschmann2021PRL}.

%%%%%%%%%%%%%%%%%%%%%%%%%%%%%%%%%%%%%%%%%%%%%%%%%%
\section{Dark vector emission in strong degenerate plasma}
\label{apd:degenerate}
%%%%%%%%%%%%%%%%%%%%%%%%%%%%%%%%%%%%%%%%%%%%%%%%%%

For the nucleon-nucleon bremsstrahlung emission of dark vectors in the NS core, we can safely assume a degenerate limit because of 
$T_c\ll \mathcal{O}(\rm MeV)$ in the core of the NS.
Let us first consider the process $n+p\to n+p+\gamma'$, the dark vector energy emission rate is given by the formula~\eqref{eq:ee2}.
Multiply the equation by one in the form~\cite{Harris2020JCAP}
\begin{equation}
  \begin{aligned}
  1 &=\int_{0}^{\infty} d p_{1} d p_{2} d p_{3} d p_{4} \delta\left(p_{1}-p_{Fn}\right) \delta\left(p_{2}-p_{Fp}\right) 
  \delta\left(p_{3}-p_{Fn}\right) \delta\left(p_{4}-p_{Fp}\right)\\
  &=\frac{1}{p_{Fp}^{2} p_{Fn}^{2}} \int d E_{1} d E_{2} d E_{3} d E_{4} E_{1}^* E_{2}^* E_{3}^* E_{4}^*
  \delta\left(p_{1}-p_{Fn}\right) \delta\left(p_{2}-p_{Fp}\right) \delta\left(p_{3}-p_{Fn}\right) \delta\left(p_{4}-p_{Fp}\right),
\end{aligned}
\end{equation}
where $p_{k}\equiv |\mathbf{p}_k|$ and $p_{Fn}$ and $p_{Fp}$ are the Fermi momenta for neutron and proton, respectively, and $p_kdp_k=E_k^*dE_k^*$ has been used. The squared matrix can be expanded as
\begin{equation}
  |\mathcal{M}|^2=a(\mathbf{k},\mathbf{l})\omega^{-\xi}.
\end{equation}
where $\xi=2$ for matrix~\eqref{eq:SMpp} and \eqref{eq:SMpn}.
% where $\mathbf{k}=\mathbf{p}_2-\mathbf{p}_4$ and $\mathbf{l}=\mathbf{p}_2-\mathbf{p}_3$. 
% For the process $p+n\to p+n+A'$, $N=2$. 
The energy emission rate~\eqref{eq:ee2} can be written as 
\begin{equation}
  Q_{\gamma'}=\frac{S}{2^{14}\pi^{10}p_{Fp}^2p_{Fn}^2}AI,
\end{equation}
where the energy integral is
\begin{equation}
  \label{eq:eint}
  I=\int d\omega \omega^{2-\xi}\int dE_{1}dE_{2}dE_{3}dE_{4}\delta\left(E_{1}+E_{2}-E_{3}-E_{4}
  -\omega\right)f_{1}f_{2}\left(1-f_{3}\right)\left(1-f_{4}\right),
\end{equation}
after using $E_{1}+E_{2}-E_{3}-E_{4}=E_{1}^{*}+E_{2}^{*}-E_{3}^{*}-E_{4}^{*}$.
In the strong degeneracy limit $\mu_j/T\to\infty$, which is a good approximation for the processes taking place in the NS core,
the energy integral is given by~\cite{Iwamoto2001PRD}
\begin{eqnarray}
  \label{eq:eneint}
  I=\frac{T^{6-\xi}}{6}\int_0^{\infty}dx\frac{x^{3-\xi}(x^2+4\pi^2)}{e^{x}-1}=\frac{11\pi^4}{90}T^4.
\end{eqnarray}

The angular integral is given by
\begin{eqnarray}
  \label{eq:Ai}
  A&=&\int d^{3}\mathbf{p}_{1}d^{3}\mathbf{p}_{2}d^{3}\mathbf{p}_{3}d^{3}\mathbf{p}_{4}\delta^{3}\left(\mathbf{p}_{1}+\mathbf{p}_{2}-\mathbf{p}_{3}-\mathbf{p}_{4}\right) \nonumber\\
  &&\times\delta\left(p_{1}-p_{Fn}\right) \delta\left(p_{2}-p_{Fp}\right) \delta\left(p_{3}-p_{Fn}\right)\delta\left(p_{4}-p_{Fp}\right)a(\mathbf{k},\mathbf{l}).
\end{eqnarray}
The dark vector momentum $\mathbf{p}_{A'}$ has been neglected in the momentum-conserving $\delta$-function.
Since the squared matrix element is in general a function of the momentum transfer $\mathbf{k}$ and $\mathbf{l}$, 
we can convert the integral to $\mathbf{k}$ and $\mathbf{l}$ by inserting the unity~\cite{Fortin2021}
\begin{equation}
  1=\int d^{3}\mathbf{k}d^{3}\mathbf{l}\delta^{3}\left(\mathbf{k}-\mathbf{p}_{2}+\mathbf{p}_{4}\right) 
  \delta^{3}\left(\mathbf{l}-\mathbf{p}_{2}+\mathbf{p}_{3}\right)
\end{equation}
into the right-hand side of Eq.~\eqref{eq:Ai}. We can eliminate $\mathbf{p}_{2}$, $\mathbf{p}_{3}$, and $\mathbf{p}_4$ one by one with the aid of the
three 3-momentum-conserving $\delta$-functions.  Then
\begin{eqnarray}
  \label{eq:int1}
  A&=&\int d^{3}\mathbf{p}d^{3}\mathbf{k}d^{3}\mathbf{l}\delta\left(|\mathbf{p}|-p_{Fn}\right)\delta\left(|\mathbf{p}+\mathbf{k}+\mathbf{l}|-p_{Fp}\right) \nonumber\\
  &&\times\delta\left(|\mathbf{p}+\mathbf{k}|-p_{Fn}\right) \delta\left(|\mathbf{p}+\mathbf{l}|-p_{Fp}\right)a(\mathbf{k},\mathbf{l}),
\end{eqnarray}
where we have relabeled $\mathbf{p}_1$ as $\mathbf{p}$.
In order to evaluate the integration, we choose the spherical coordinates to expand the three vectors as
\begin{eqnarray}
  \mathbf{p}=p(0,~0,~1),
  ~
  \mathbf{k}=k(\sin\theta_k,~0,~\cos\theta_k),
  ~
  \mathbf{l}=l(\sin\theta_l\cos\phi_l,~\sin\theta_l\sin\phi_l,~\cos\theta_l).
\end{eqnarray}
Namely, $\mathbf{p}$ lies along the $z$ axis, $\mathbf{k}$ lies in the $x-z$ plane, and $\mathbf{l}$ points to an arbitrarily direction.
Then we have
\begin{eqnarray}
  |\mathbf{p}+\mathbf{k}|^2&=&k^2+2kp\cos\theta_k+p^2,\\
  |\mathbf{p}+\mathbf{l}|^2&=&l^2+2lp\cos\theta_l+p^2,\\
  |\mathbf{p}+\mathbf{k}+\mathbf{l}|^2&=&p_{Fp}^2+2kl\sin\theta_k\sin\theta_l\cos\phi_l+2kl\cos\theta_k\cos\theta_l.
\end{eqnarray}
The $\delta$-functions in Eq.~\eqref{eq:int1} indicate $|\mathbf{p}|=p=p_{Fn}$ and
\begin{equation}
  \mathbf{k}\cdot\mathbf{l}=kl\sin\theta_k\sin\theta_l\cos\phi_l+kl\cos\theta_k\cos\theta_l=0.
\end{equation}
We see that the term proportional to $\mathbf{k}\cdot\mathbf{l}$ does not contribute. This is the consequence of strong degeneracy limit.

The $\delta$-functions in Eq.~\eqref{eq:int1} can be written as 
\begin{eqnarray}
  \delta(|\mathbf{p}+\mathbf{k}|-p_{Fn})&=&\frac{\delta(x_k-x_k^0)}{k},\\
  \delta(|\mathbf{p}+\mathbf{l}|-p_{Fp})&=&\frac{p_{Fp}\delta(x_l-x_l^0)}{lp_{Fn}},\\
  \delta(|\mathbf{p}+\mathbf{k}+\mathbf{l}|-p_{Fp})&=&\frac{p_{Fp}\delta(x_{\phi}-x_{\phi}^0)}{kl\sqrt{(1-x_k^2)(1-x_l^2)}},
\end{eqnarray}
with the definitions $x_k=\cos\theta_k$, $x_l=\cos\theta_l$, and $x_{\phi}=\cos\phi_l$, and
\begin{eqnarray}
  x_{k}^0=-\frac{k}{2p_{Fn}},~
  x_{l}^0=\frac{p_{Fp}^2-p_{Fn}^2-l^2}{2lp_{Fn}},~
  x_{\phi}^0=\frac{-x_kx_l}{\sqrt{(1-x_k^2)(1-x_l^2)}}.
\end{eqnarray} 
Using the above relations, the integration~\eqref{eq:int1} can be simplified as
\begin{eqnarray}
  A=32\pi^2p_{Fp}^2p_{Fn}^2\int dk\int dl \frac{la(\mathbf{k},\mathbf{l})}{\sqrt{4l^2p_{Fn}^2-l^2k^2-(p_{Fp}^2-p_{Fn}^2-l^2)^2}},
\end{eqnarray}
with the constraints on the parameters $-1\leq x_{k},~x_{l},~x_{\phi}\leq 1$ and $k,~l>0$.

Similarly, for the process $n+n\to n+n+A'$ the angular integral is given by
\begin{eqnarray}
  A=32\pi^2p_{Fn}^4\int dk\int dl \frac{a(\mathbf{k},\mathbf{l})}{\sqrt{4p_{Fn}^2-k^2-l^2}}.
\end{eqnarray}

\section{Remarks on OPE approximation}
\label{apd:OPE}

In order to realistically determine the production rate of new bosons ({\it e.g.} axion or dark vector) from nucleon-nucleon bremsstrahlung
processes, the simplified treatment based on one-pion exchange and the use of the Born approximation for the nucleon-nucleon interaction 
was originally introduced in Refs.~\cite{Friman1979APJ,Iwamoto1984PRL,Brinkmann1988PRD,Turner1989PRD,Carena1989PRD}. 
It has been realized in the literature~\cite{Hanhart2001PLB,Schwenk2004PLB,Rrapaj2016PRC,Chang2018JHEP,Carenza2019JCAP} that such a simplified 
method ignores some relevant effects such as the multiple nucleon scatterings, and leads to a larger emission rate and thus a stronger 
constraint on the coupling. 

One way to go beyond the OPE approximation has been performed in Refs.~\cite{Hanhart2001PLB,Schwenk2004PLB} by including a nonperturbative 
treatment of the two-nucleon scattering in the soft-radiation approximation. 
It was found that for the range of conditions encountered in a NS, this treatment results in an approximate modification of the axion emissivity 
by a factor of $1/4$ when compared with the OPE results~\cite{Beznogov2018PRC}. 
Using the soft-radiation approximation, Ref.~\cite{Rrapaj2016PRC} calculated the dark vectors emissivity from a supernova and found a factor 
of 10 decrease in the emission rate. 
Reference~\cite{Carenza2019JCAP} refined the calculation based on the OPE approximation by systematically taking into account the effects that had been neglected previously, 
including the contribution of the two-pions exchange, effective in-medium nucleon masses and multiple nucleon scatterings.
They found that the axion emissivity from a supernova was reduced by $\sim 10$ with respect to the basic OPE calculation.
From these results we observe that the reduction of emissivity depends on the condition of matter.  For the highly compressed matter in a NS,
the emissivity is reduced by a factor of 4 with respect to the OPE approximation, while the factor can be up to 10 for the dilute plasma in a supernova.

Further improvement of the soft-radiation approximation can be made by including the many-body effects, 
such as the Landau-Pomeranchuk-Migdal (LPM) suppression~\cite{Landau1953,Migdal1956PR} owing to multiple scatterings in the medium.
It has been shown in Ref.~\cite{Dalen1003PRC} that the LPM suppression of soft radiations increases with temperature and
becomes relevant for $T\sim 5$~MeV, above which the energy of the emitted boson is less than the nucleon decay width.
Thus, the LPM effect is of particular importance for radiation in a supernova, but it becomes insignificant for the isolated NSs with 
a core temperature $T\sim 10$~keV.

\end{document}